\documentclass[12pt]{article}
\usepackage{a4wide}
\usepackage{epsfig}
\newcommand{\eps}{\varepsilon}
\newcommand{\slk}{/\kern-6pt k}
\newcommand{\sll}{/\kern-4pt l}
\newcommand{\slp}{p\kern-5pt/}
\newcommand{\slq}{q\kern-5.5pt/}

\newcommand{\Tr}{\mathop{\rm Tr}\nolimits}
\newcommand{\oone}{\hbox{$1\kern-2.5pt\hbox{\rm l}$}}
\newcommand{\ssigma}{\hbox{$\kern2.5pt\vrule height4pt\kern-2.5pt\sigma$}}

\newcommand{\GeV}{{\rm\,GeV}}
\newcommand\pfrac[2]{\left(\frac{#1}{#2}\right)}
\newcommand{\Li}{{\rm Li}}
\newcommand{\imag}{\mathop{\rm Im}\nolimits}
\newcommand{\real}{\mathop{\rm Re}\nolimits}
\newcommand{\sla}{{\sqrt\lambda}}
\newcommand{\IP}{\mbox{I}\!\mbox{P}}

\begin{document}

\thispagestyle{empty} 
\begin{flushright}
MZ-TH/12-55\\
December 2012
\end{flushright}
\vspace{0.1cm}

\begin{center}
{\Large\bf Fully analytical $O(\alpha_s)$ results for on-shell and off-shell 
\\[.2cm]  
polarized $W$-boson decays into massive quark pairs}\\[1.3cm]
{\large S.~Groote$^{1,2}$, J.G.~K\"orner$^2$ and P.~Tuvike$^1$}\\[1cm]
$^1$ Loodus- ja Tehnoloogiateaduskond, F\"u\"usika Instituut,\\[.2cm]
  Tartu \"Ulikool, T\"ahe 4, 51010 Tartu, Estonia\\[7pt]
$^2$PRISMA Cluster of Excellence, Institut f\"ur Physik,
  Johannes-Gutenberg-Universit\"at,\\[.2cm]
  Staudinger Weg 7, 55099 Mainz, Germany
\end{center}

\begin{abstract}\noindent
We provide analytical $O(\alpha_s)$ results for the three polarized decay
structure functions $H_{++},\,H_{00}$ and $H_{--}$ that describe the decay of
a polarized $W$ boson into massive quark--antiquark pairs. As an application
we consider the decay $t\to b+W^+$ involving the helicity fractions
$\rho_{mm}$ of the $W^+$ boson followed by the polarized decay
$\hbox{$W^+(\uparrow)$}\to q_1\bar{q}_2$ described by the polarized decay
structure functions $H_{mm}$. We thereby determine the $O(\alpha_s)$ polar
angle decay distribution of the cascade decay process
$t\to b+W^+(\to q_1\bar{q}_2)$. As a second example we analyze quark mass and
off-shell effects in the cascade decays
$H\to W^{-}+W^{\ast +}(\to q_1\bar{q}_2)$ and $H\to Z+Z^{\ast}(\to q\bar{q})$. 
For the decays $H\to W^{-}+W^{\ast +}(\to c\bar b)$ and 
$H\to Z+Z^{\ast}(\to b\bar{b})$ we find substantial deviations from the
mass-zero approximation in particular in the vicinity of the threshold region.
\end{abstract}

\newpage

\section{Introduction}
The polarization of $W^\pm$ bosons produced in electroweak production 
processes is in general highly nontrivial. Therefore, the $W^\pm$ bosons
produced e.g.\ in $pp(p\bar p)\to W^{\pm}+X$, $pp(p\bar p)\to W^+W^-+X$, 
$e^+e^-\to W^+W^{-},\,W^+W^{-}+X$ and $t\to b+W^+$ in general have a highly
nontrivial polarization density matrix. Because of this, there is a rich
phenomenology of polarization effects in $W$ production and decay to be
explored in present and future experiments. For example, one would want to
compare the results of polarization measurements with the predictions of the
Standard Model (SM) or models beyond the SM.

The polarization of the $W^\pm$ bosons can be probed by decay correlations
involving the decay products of the polarized $W^\pm$ boson. Using such decay
correlations, first measurements of the $W^{\pm}$ polarization in 
$pp\to W^{\pm}+X$ were reported by the CMS
Collaboration~\cite{Chatrchyan:2011ig} and the ATLAS
Collaboration~\cite{ATLAS:2012au}. Measurements of the $W^\pm$ polarization in
$e^+e^-\to W^+W^-$ were published in
Refs.~\cite{Abbiendi:2003wv,Achard:2002bv}. Finally, results of
$W^+$-polarization measurements in $t\to b+W^+$ were presented e.g.\ in
Refs.~\cite{Acosta:2004mb,Abulencia:2006ei,Aaltonen:2008ei,Abazov:2007ve,
Aaltonen:2012tk,:2012hy}. Ref.~\cite{Stirling:2012zt} provides a survey
of SM expectations for the polarization of $W$ bosons in various production
channels at the LHC. 

In the SM the $W^\pm$ boson decays into quark or lepton pairs. For unpolarized
$W^\pm$-boson decays the NLO QCD and electroweak corrections to quark and
lepton pair production, resp., have been given in
Ref.~\cite{Denner:1990tx,Denner:1991kt}. The radiative corrections in
Ref.~\cite{Denner:1990tx,Denner:1991kt} include also quark and lepton-mass 
effects. To our knowledge the radiative corrections to polarized $W^\pm$-boson
decays including lepton and quark mass effects have not been done up to now. 

This paper is devoted to the evaluation of the NLO QCD corrections to the
decays of polarized $W^\pm$ bosons into massive quark--antiquark pairs
$W^{\pm}(\uparrow)\to q_1\,\bar q_2$ where the diagonal spin density
matrix elements of the $W^\pm$ boson can be probed through the polar angle 
decay distribution of the final-state quark pair. We augment our results such 
that they can also be applied to the decay of polarized $Z$ decays into 
massive quark pairs. In order to provide quick access to the importance of
quark mass effects in the decays of the $W^\pm$ and $Z$ bosons we have
provided a $O(m_{q_i}^2/m_W^2)$ quark mass expansions of our analytical
results in a separate paper~\cite{Groote:2012xr}. In a sequel to the present
paper we shall  calculate the corresponding NLO electroweak corrections to the
polarized decay $W^+(\uparrow)\to\ell^+\,\nu_\ell$~\cite{electroweak11}.

In the limit $m_{q_i}=:m_i\to 0$ our results reduce to rather simple forms
which agree with previous NLO QCD results extracted from the corresponding
calculation of
$(\gamma^*,Z)(\uparrow)\to q\bar q$~\cite{Groote:1995yc,Groote:1995ky,
Groote:1996nc, Groote:2008ux,Groote:2009zk}. Quark mass effects are 
non-negligible even for on-shell $W$ bosons with $q^2=m_W^2$ for the polarized
decay  $W^+(\uparrow)\to c\bar b$ but become even more important for
lower values of $q^2$ as for the decays of off-shell $W^{\ast\pm}$ and
$Z^{\ast}$ bosons as they appear e.g.\ in the recently observed discovery
channels $H\to W^\pm W^{\ast\mp}$ and $H\to ZZ^\ast$ of a $126\GeV$ Higgs
boson~\cite{:2012gk,:2012gu}. Similarly one needs to retain mass effects in
the calculation of current--current correlators and their corresponding
spectral functions which are needed for all values of $q^2$. Since there have
been claims and counterclaims in the literature as to the correctness of known
results on radiative corrections to scalar (pseudoscalar) and vector
(axial-vector) current--current spectral functions, we have compared our
unpolarized results with previously published spectral function results.

As an illustration of our general decay analysis we consider the cascade decay
process $t\to b +W^+$ followed by $W^+\to q_1\,\bar q_2$ where the (helicity
frame) diagonal density matrix elements of the $W^+$ boson resulting from the
decay process $t\to b+W^+$ have been well studied in the literature. We thus
provide results on the angular decay distribution for the sequential cascade
decay $t\to b+W^+(\to q_1\,\bar q_2)$ for which we discuss NLO QCD radiative
corrections in the production process $t\to b+W^+(\uparrow)$ and in the decay
process $W^+(\uparrow)\to q_1\,\bar q_2$. As a second example of much topical
interest we take the cascade decay processes
$H\to W^-+W^{\ast+}(\to q_1\bar q_2)$ and $H\to Z+Z^\ast(\to q\bar q)$ where
we discuss quark mass and $W^\ast$ and $Z^\ast$ off-shell effects on rates
and on angular decay distributions.  

We also briefly comment on the nondiagonal density matrix elements of the
$W^\pm$ boson which can be probed by azimuthal correlations in the angular
decay distribution. A measurement of the azimuthal correlations requires the
existence of a preferred transverse direction which would be provided e.g.\ by
the transverse polarization direction of the polarized top quark in the decay
$t(\uparrow)\to b+W^+(\to q_1\,\bar q_2)$. In a similar vein a transverse
direction can be defined in the large-$p_T$ $W$-boson production in the
process $pp(p\bar p)\to W+X$.

\section{Born-term results}
\begin{figure}\begin{center}
\epsfig{figure=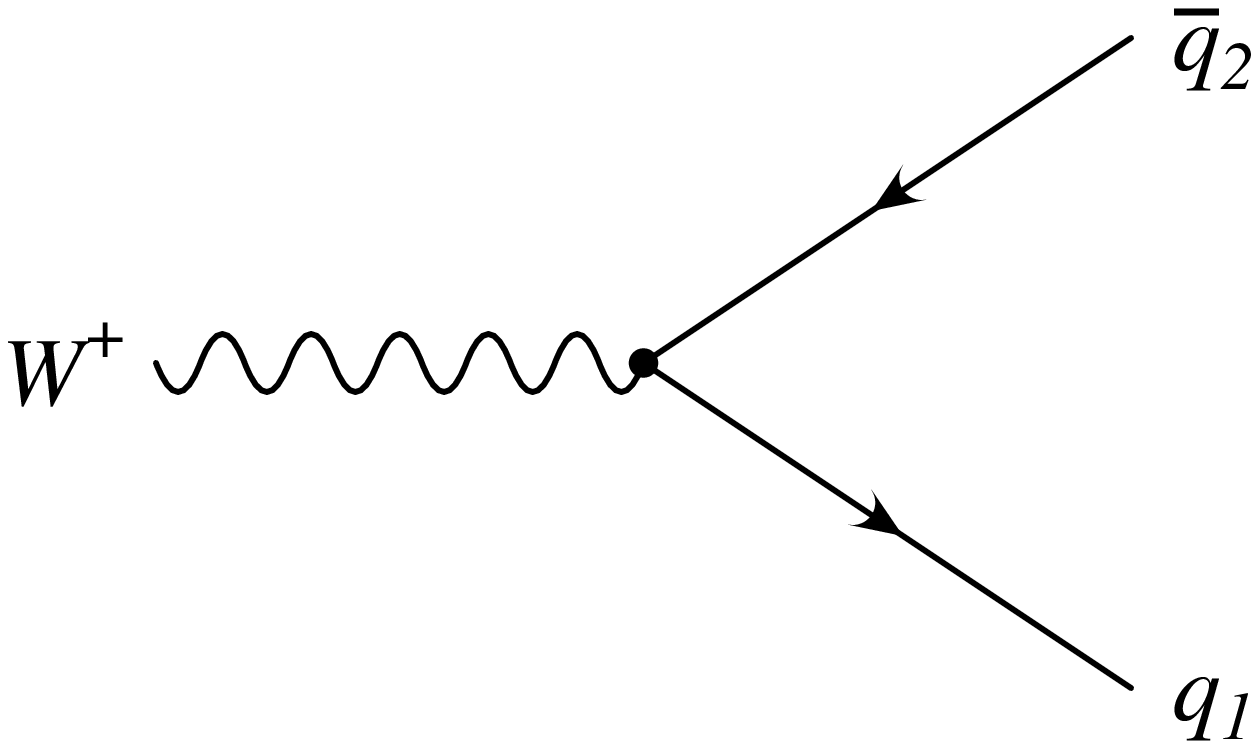, scale=0.5}\qquad
\epsfig{figure=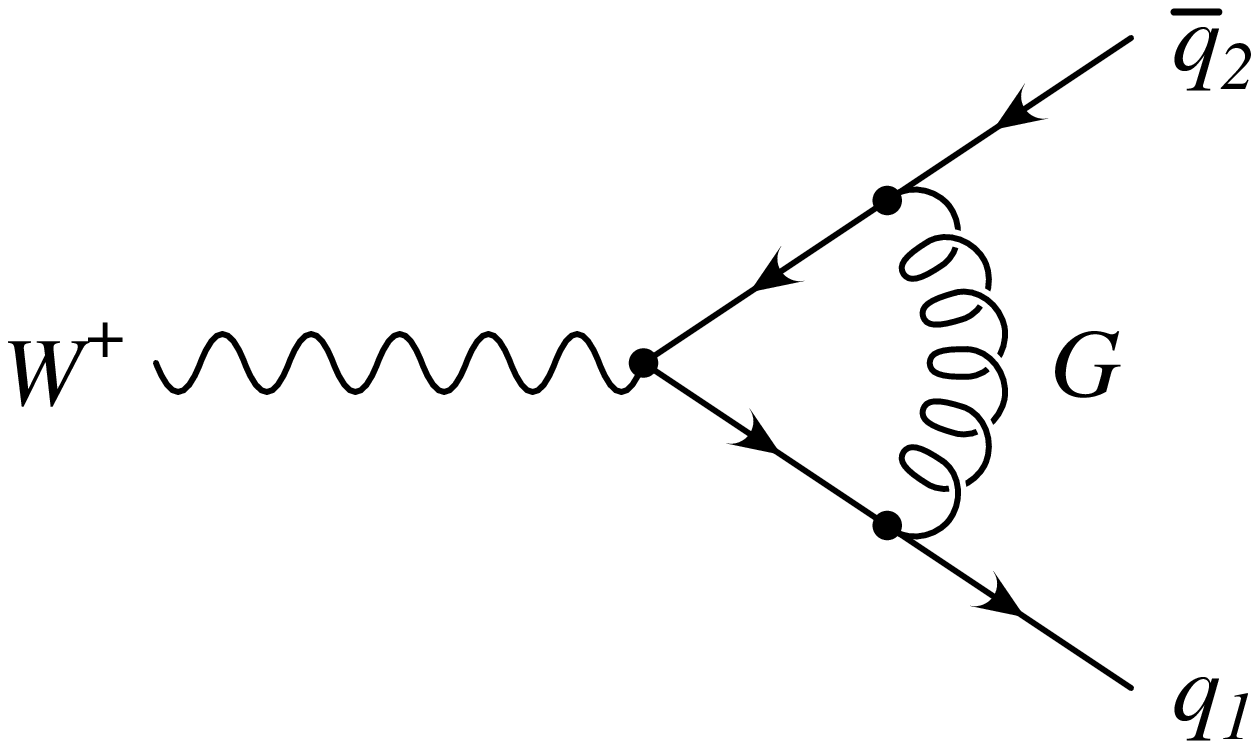, scale=0.5}\\[12pt]
(a)\kern164pt(b)
\end{center}
\caption{\label{wlagu03}Feynman diagrams for (a) the Born-term contribution 
and (b) the one-loop QCD contribution to the decay process
$W^+\to q_1\,\bar q_2$}
\end{figure}

Let us consider the quark--antiquark decay of the SM gauge boson $W^+$
\begin{equation}
W^+(q)\to q_1(p_1)\,\bar q_2(p_2)
\end{equation}
as depicted in Fig.~\ref{wlagu03}.
The LO Born-term amplitude is given by 
\begin{equation}\label{loampl}
{\cal M}({\it Born\/})={\cal M}^\mu({\it Born\/})\varepsilon_\mu(q)
  =-i\frac{g_w}{\sqrt2}V_{ij}\,\,\bar u_1(p_1)\gamma^\mu\frac{1-\gamma_5}2
  v_2(p_2)\,\varepsilon_\mu(q),
\end{equation}
where $g_w$ is the electroweak coupling constant and the $V_{ij}$ are
Kobayashi--Maskawa matrix elements ($q_1=i$; $q_2=j$). We define a reduced
matrix element $\widetilde{{\cal M}}^\mu$ by splitting off the common coupling
factor $-ig_WV_{ij}/\sqrt2$ and the factor $1/2$ from the chiral projector.
The reduced Born-term tensor reads
\begin{eqnarray}
H^{\mu\nu}({\it Born\/})&=&N_c\sum_{\rm quark\ spins}
 \widetilde{{\cal M}}^\mu({\it Born\/})
\widetilde{{\cal M}}^{\dagger\nu}({\it Born\/})\nonumber\\
  &=&N_c\,\Tr\left((\slp_1+m_1)\gamma^\mu(1-\gamma_5)(\slp_2-m_2)\gamma^\nu
  (1-\gamma_5)\right)\nonumber\\
  &=&8N_c\left(p_1^\mu p_2^\nu+p_1^\nu p_2^\mu-p_1p_2\,g^{\mu\nu}
  +i\eps^{\mu\nu\alpha\beta}p_{1\alpha}p_{2\beta}\right).
\end{eqnarray}
The Born-term amplitude (\ref{loampl}) leads to the LO decay width
for an on-shell $W^+$ boson with $q^2=m_W^2$
($g_w^2=4\pi\alpha/\sin^2\theta_W$, $\mu_i=m_i^2/q^2$),
\begin{eqnarray}
\label{lowidth}
\Gamma({\it Born\/})&=&\frac13\,\frac1{8\pi}\,\frac{|\vec{p}|}{m_W^2}
\frac{g_w^2}2|V_{ij}|^2N_c H_{\mu\nu}({\it Born\/})
  \left(-g_{\mu\nu}+\frac{q_{\mu}q_{\nu}}{m_W^2}\right)\nonumber\\
  &=&\frac{m_W}{96\pi}g_W^2|V_{ij}|^2N_c\,\sla
  \left(2-\mu_1-\mu_2-(\mu_1-\mu_2)^2\right),
\end{eqnarray}
where $|\vec p\,|=m_W\sla/2$, and where $\lambda$ is the value of the
K\"all\'en function for the decay process,
\begin{equation}
\lambda=\lambda(1,\mu_1,\mu_2)=1+\mu_1^2+\mu_2^2-2\mu_1-2\mu_2-2\mu_1\mu_2.
\end{equation}
The rate expression~(\ref{lowidth}) coincides with the Born-term result in 
Ref.~\cite{Denner:1990tx}.

The subject of this paper are the partial decays from states of the $W^+$
boson with definite $m$ quantum numbers $m=\pm1,0$, i.e.\ we are interested in
the polarized decay structure functions
\begin{equation}
\label{proj1}
H_{\pm\pm}=H_{\mu\nu}\,\varepsilon^\mu(\pm)\varepsilon^{*\nu}(\pm),
\qquad H_{00}=H_{\mu\nu}\,\varepsilon^{\mu}(0)\varepsilon^{*\nu}(0).
\end{equation}
We evaluate the polarized decay functions defined in Eq.~(\ref{proj1}) in the
rest frame of the $W^+$ boson with the $z'$ direction defined by the antiquark
$\bar q_2$.\footnote{We have chosen the antiquark direction to define the $z'$
axis in analogy to the antilepton $\ell^+$ in the decay $W^+\to\ell^+\nu_\ell$.
One can equally well choose the quark to define the $z'$ axis. The resulting
changes in the partial helicity rate functions will be discussed later on.}
The rest frame polarization vectors and momenta are thus given by
\begin{eqnarray}
\varepsilon^\mu(\pm)&=&\frac1{\sqrt2}\,\Big(0;\mp 1,-i,0\Big),\qquad
q^\alpha=\Big(m_W;0,0,0\Big),\nonumber\\
\varepsilon^\mu(0) &=&\Big(0;0,0,1\Big),\qquad\qquad 
p_2^\alpha=(E_2;0,0,|\vec{p}\,|),
\end{eqnarray}
where $E_2=m_W(1-\mu_1+\mu_2)/2$ and
$|\vec p\,|=\sqrt{q^2}\sqrt{\lambda(1,\mu_1,\mu_2)}/2$.

It proves convenient to bring the rest frame projectors
$\IP^{\mu\nu}_{\pm\pm}=\varepsilon^\mu(\pm)\varepsilon^{*\nu}(\pm)$ and
$\IP^{\mu\nu}_{00}=\varepsilon^\mu(0)\varepsilon^{*\nu}(0)$ into a
frame-independent covariant form. One has
\begin{equation}\label{proj2}
\IP^{\mu\nu}_{\pm\pm}\ =\ \frac12\left(\IP^{\mu\nu}_{U+L}-\IP^{\mu\nu}_L
  \pm\IP^{\mu\nu}_F\right),\qquad
\IP^{\mu\nu}_{00}\ =\ \IP^{\mu\nu}_L,
\end{equation}
where
\begin{eqnarray}
\IP^{\mu\nu}_{U+L}&=&-g^{\mu\nu}+\frac{q^\mu q^\nu}{q^2},\nonumber\\
\IP^{\mu\nu}_L&=&\frac{q^2}{N_P^2}
    \Big(p_2^\mu-\frac{p_2\cdot q}{q^2}q^\mu\Big)
    \Big(p_2^\nu-\frac{p_2\cdot q}{q^2}q^\nu\Big),\nonumber\\
\IP^{\mu\nu}_F&=&\frac1{N_P}i\epsilon^{\mu\nu\alpha\beta}p_{2\alpha}q_\beta,
\end{eqnarray}
and where the normalization factor $N_P$ is given by
$N_P^2=((p_2q)^2-p_2^2q^2)$. In the two-body case the normalization factor
is reduced to $N_{P}=\sqrt{q^2}\,|\vec p\,|$. The covariant form of the
projectors are particularly convenient in the NLO tree-graph calculation since
the covariantly projected integrands in the requisite phase space integrations
are Lorentz scalars and can thus be handled by the standard covariant methods.

Using either forms for the projectors~(\ref{proj1}) or~(\ref{proj2}), one
obtains
\begin{equation}\label{bornpol}
H_{\pm\pm}({\it Born\/})=4N_cq^2(1-\mu_1-\mu_2\pm\sla),\qquad
H_{00}({\it Born\/})=4N_cq^2(1-\mu_1-\mu_2-\lambda).
\end{equation}
Note that the sum $H_U=H_{++}+H_{--}$ ($U$: unpolarized transverse) and
$H_L=H_{00}$ ($L$: longitudinal) are fed only by the parity-even $VV$ and 
$AA$ current products. The difference $H_F=H_{++}-H_{--}$ ($F$:
forward--backward asymmetric) is fed by the parity-odd $V\!A$ current product.

At threshold, where $q^2\to(m_1+m_2)^2$, with $\sqrt{\mu_1}+\sqrt{\mu_2}\to 1$
and $\lambda\to 0$, one has
$H_{--}({\it Born\/})=H_{00}({\it Born\/})=H_{++}({\it Born\/})=8N_cm_1m_2$.
All three partial helicity rates are equal to one another at threshold. This
can be understood from the fact that, at threshold, only the vector
current-induced $LS$ amplitude $(LS)=(01)$ survives. This leads to the
equality of the partial helicity rates using simple Clebsch--Gordan algebra.
As we shall see in the next section, at threshold one loses the analyzing
power of the two-fermion decay mode, i.e.\ the angular decay distribution
becomes flat at threshold irrespective of the polarization of the $W^+$ boson. 

In the massless quark limit $\mu_1=\mu_2=0$ one has
$H_{++}({\it Born\/})=8N_cq^2\neq 0$ and
$H_{00}({\it Born\/})=H_{--}({\it Born\/})=0$ as expected from the left-chiral
nature of the SM current (\ref{loampl}). The finite mass corrections to the LO
helicity structure functions are of $O(\mu_i)$ for $H_{++}$ and $H_{00}$,
i.e.\ $H_{++}({\it Born\/})=8N_cq^2(1-\mu_1-\mu_2+\ldots)$ and
$H_{00}({\it Born\/})=4N_cq^2(\mu_1+\mu_2+\ldots)$, and of $O(\mu_i^2)$ for
$H_{--}$, i.e.\ $H_{--}({\it Born\/})=4N_cq^2(\mu_1\mu_2+\ldots)$. For the sum
of the three polarized decay functions denoted by $H_{U+L}$ one obtains
\begin{eqnarray}\label{hupBorn}
H_{U+L}({\it Born\/})&=&H_{--}({\it Born\/})+H_{00}({\it Born\/})
  +H_{++}({\it Born\/})\nonumber\\
  &=&12N_cq^2\left(1-\mu_1-\mu_2-\lambda/3\right).
\end{eqnarray}

For the sake of completeness we also define a scalar structure function
$H_{tt}$ through $H_{tt}=H_{\mu\nu}\,\varepsilon^\mu(t)\varepsilon^{*\nu}(t)$
where $\varepsilon^\mu(t)=(1;0,0,0)$ is the rest-frame time-component (scalar)
polarization vector of the off-shell $W^+$ boson. The corresponding
covariant projector on the scalar structure function reads
\begin{equation}
\IP^{\mu\nu}_S=\frac{q^\mu q^\nu}{q^2}.
\end{equation}
For the LO scalar structure function one obtains 
\begin{equation}
H_{tt}({\it Born\/})=H_S({\it Born\/})
  =4N_cq^2\left(1-\mu_1-\mu_2-\lambda\right).
\end{equation}
Note that, at the Born-term level, one has
$H_{tt}({\it Born\/})=H_{00}({\it Born\/})$. $H_{tt}({\it Born\/})$ vanishes
for zero quark masses as expected from current conservation in the mass-zero
limit.

The scalar--longitudinal interference term needed later on is projected by
\begin{equation}
\IP^{\mu\nu}_{t0}=\frac1{N_P}q^\mu\,
\Big(p_2^\nu-\frac{p_2\cdot q}{q^2}q^\nu\Big),
\end{equation}
such that
\begin{equation}
H_{0t}({\it Born\/})=H_{t0}({\it Born\/})
  =-4N_cq^2\left(\mu_1-\mu_2\right)\sla.
\end{equation}

\section{Angular decay distribution and\\
  the cascade decay $t\to b+W^+(\to q_i\,\bar q_j)$}
Consider the rest frame decay of a polarized $W^+$ with the diagonal spin
density matrix elements ($\rho_{++},\rho_{00},\rho_{--}$) given in an unprimed
coordinate system $(x,y,z)$. Then rotate the coordinate system $(x,y,z)$
around the $y$ axis by an angle $\theta$ to a primed coordinate system
$(x',y,z')$. Under this rotation the diagonal density matrix elements
transform according to
$\rho'_{m'm'}(\theta)=\rho_{mm}\,d^1_{mm'}(\theta)d^1_{mm'}(\theta)$.
The angular decay distribution is then determined by the product of the decay
probability $H_{m'm'}$ for the decay $W^+(m')\to q_1\,\bar q_2$ and the
relevant diagonal elements of the spin density matrix elements
$\rho'_{m'm'}(\theta)$, all evaluated in the primed system.

While a decay analysis in the $W^+$ rest system is the optimal choice to probe
the density matrix elements of the $W^+$ boson, the polarization of the
$W^+$ boson can also be detected in other coordinate systems. As an example
take the cascade decay $t\to b+W^+(\to\ell^+\,\nu_\ell)$. When analyzed in the
top quark rest system, the polarization of the $W^+$ will affect the energy
spectrum of the final lepton, i.e.\ leptons from $\rho_{--}$ will be more
energetic than those from $\rho_{++}$.

Returning to the analysis in the $W^+$ rest frame we mention that the choice
of the $z$ and $z'$ axes is a matter of convention and convenience and may be
dictated by the physics at hand. For example, in the process
$pp(p\bar p)\to W^++X$ followed by $W^+ \to\ell^+\nu$ several unprimed
rest frame coordinate systems have been discussed in the literature
(Collins--Soper frame, recoil frame, target frame, beam frame) whereas the
$z'$ direction is conventionally fixed by the lepton direction.\footnote{NLO
results on $W$ polarization effects in $p\bar p\to W+X$ can be found in
Refs.~\cite{Mirkes:1990vn,Mirkes:1992hu}.} 

In the example discussed further on ($t\to b+W^+(\to q_1\,\bar q_2)$) the $z$
direction is fixed by the momentum direction of the $W^+$ in the top quark
rest system (helicity system), and the $z'$ direction is determined by the
momentum direction of the antiquark $\bar q_2$.

It is convenient to work in terms of normalized spin density matrix elements
defined by $\hat\rho_{mm}=\rho_{mm}/\sum_{m'}\rho_{m'm'}$ with
$\hat\rho_{++}+\hat\rho_{00}+\hat\rho_{--}=1$ and normalized decay functions
given by $\hat H_{mm}=H_{mm}/\sum_{m'}H_{m'm'}$ such that
$\hat H_{++}+\hat H_{00}+\hat H_{--}=1$. According to what was said before, the
normalized decay distribution is given by
\begin{eqnarray}\label{Wtheta}
\widehat W(\theta)&=&\frac32\quad
\sum_{m,m'=0,\pm}\hat\rho_{mm}\,d^1_{mm'}(\theta)
  \,d^1_{mm'}(\theta)\,\hat H_{m'm'}\nonumber\\
  &=&\frac38(1+\cos^2\theta)\,(\hat\rho_{++}+\hat\rho_{--})\,
  (\hat H_{++}+\hat H_{--})
  +\frac34\cos\theta\,(\hat\rho_{++}-\hat\rho_{--})\,
  (\hat H_{++}-\hat H_{--})\nonumber\\&&\strut
  +\frac34\sin^2\theta\,(\hat\rho_{++}\,\hat H_{00}+\hat\rho_{00}\,\hat H_{++}
  +\hat\rho_{00}\,\hat H_{--}+\hat\rho_{--}\,\hat H_{00})
  +\frac32\cos^2\theta\,\hat\rho_{00}\,\hat H_{00}\nonumber\\ 
  &=&\frac38\cos^2\theta\,(\hat\rho_{++}-2\hat\rho_{00}+\hat\rho_{--})
  (\hat H_{++}-2\hat H_{00}+\hat H_{--})\nonumber\\&&\strut
  +\frac34\cos\theta(\hat\rho_{++}-\hat\rho_{--})\,(\hat H_{++}-\hat H_{--})
  \nonumber\\&&\strut
  +\frac38\Big((\hat\rho_{++}+2\hat\rho_{00}+\hat\rho_{--})
  (\hat H_{++}+2\hat H_{00}+\hat H_{--})-4\hat\rho_{00}\hat H_{00}\Big).
\end{eqnarray}
The distribution~(\ref{Wtheta}) is a second-degree polynomial in $\cos\theta$
and therefore has the form of a parabola. Integrating over $\cos\theta$ one
obtains 
\begin{equation}
\label{IWtheta}
\int\widehat W(\theta)\,d\cos\theta=1.
\end{equation}
For unpolarized $W^+$ decay one has
$\hat\rho_{--}=\hat\rho_{00}=\hat\rho_{++}=1/3$ which results in a flat decay
distribution $\widehat W(\theta)=1/2$. Similarly, one obtains a flat decay
distribution at threshold where
$ \hat H_{--}= \hat H_{00}=\hat H_{++}=1/3$, i.e.\
$\widehat W(\theta)\propto(\hat\rho_{--}+\hat\rho_{00}+\hat\rho_{++})/2=1/2$ 
irrespective of the polarization of the $W$ boson.

In the zero quark mass limit and to leading order in $\alpha_s$ (where
$\hat H_{++}({\it Born\/})=1$ and
$\hat H_{00}({\it Born\/})=\hat H_{--}({\it Born\/})=0$) the angular decay
distribution~(\ref{Wtheta}) reduces to
\begin{equation}\label{Wtheta2}
\widehat W(\theta)\ =\ \frac38(1+\cos\theta)^2\,\hat\rho_{++}
  +\frac38(1-\cos\theta)^2\,\hat\rho_{--}
  +\frac34\sin^2\theta\,\hat\rho_{00},
\end{equation}
a form quite familiar from the analysis of the cascade decay
$t\to b+W^+(\to \nu_\mu\,\mu^+)$~\cite{Acosta:2004mb,Abulencia:2006ei,%
Aaltonen:2008ei,Abazov:2007ve,Aaltonen:2012tk}.

Let us now turn to the $\alpha_s$ corrections to the polarized decay
functions $H_{mm}$ where we include quark mass effects. Surprisingly it turns
out that the quark mass corrections to the leading NLO term set in linearly
and carry rather large coefficients. This has to be contrasted with the LO and
the NLO unpolarized decay term where the mass corrections set in
quadratically. In fact, expanding the $O(\alpha_s)$ polarized decay functions
$H_{mm}$ listed in Sec.~7 up to $O(\sqrt{\mu_i})$, one obtains (see also
Ref.~\cite{Groote:2012xr} where the expansion is carried out to $O(\mu_i)$)
\begin{eqnarray}
H_{++}&=&8N_cq^2\bigg[\,1+\frac{\alpha_s}{6\pi}
  \Big(1+(\pi^2+16)\sqrt{\mu_2}\Big)+\ldots\bigg],\nonumber\\
H_{00}\,\,\,&=&8N_cq^2\bigg[\,0+\frac{\alpha_s}{6\pi}
  \Big(4-2\pi^2\sqrt{\mu_2}\Big)+\ldots\bigg],\nonumber\\
H_{--}&=&8N_cq^2\bigg[\,0+\frac{\alpha_s}{6\pi}
  \Big(1+(\pi^2-16)\sqrt{\mu_2}\Big)+\ldots\bigg]. 
\end{eqnarray}
The NLO linear mass corrections are proportional to the antiquark mass $m_2$
and are thus maximally asymmetric in the quark masses.\footnote{When one
chooses the $z'$ direction along the quark direction (called system~I in
Ref.~\cite{Groote:2012xr}), the linear mass corrections are proportional to
the quark mass $m_1$. As discussed in Ref.~\cite{Groote:2012xr}, the polarized
decay functions $H^{I}_{mm}$ in this system are obtained from the present
results by the substitution $H^{II}_{\pm\pm}(1,2)\to H^{I}_{\mp\mp}(2,1)$ and
$H^{II}_{00}(1,2)\to H^{I}_{00}(2,1)$ where, using the notation of
Ref.~\cite{Groote:2012xr}, the polarized decay functions described in this
paper are denoted by $H^{II}_{mm}(1,2)$.} It is apparent that the NLO linear
mass terms cancel in the sum $H_{++}+H_{00}+H_{--}$. We mention that the
leading order $O(\mu_i^0)$ $\alpha_s$ contributions can also be extracted
from the corresponding calculation of
$(\gamma^*,Z)(\uparrow)\to q\bar q$ in Refs.~\cite{Groote:1995yc,%
Groote:1995ky,Groote:1996nc,Groote:2008ux,Groote:2009zk} when the quark masses
are set to zero in these calculations. As concerns the leading order
$\alpha_s$ contributions, the largest contribution occurs for $H_{00}$ and 
amounts to $2\alpha_s/(3\pi)=2.5\,\%$ with $\alpha_s(m_W^2)=0.117$. The
$\alpha_s$ corrections can be seen to sum up to 
$H_{++}+H_{00}+H_{--}\sim(1+\alpha_s/\pi)$, a result which is well familiar 
from $e^+e^-$ annihilation into mass-zero quark pairs.

The NLO linear mass corrections have rather large coefficients. For example
for $W^+\to c\bar b$ and for the polarized structure function $H_{++}$, which
is the only polarized structure function with a sizeable LO contribution, the
linear mass correction amounts to $155\,\%$ (with $m_b=4.8\GeV$ and
$m_W=80.399\GeV$). However, the large mass corrections are tempered when one
calculates the normalized decay functions $\hat H_{mm}$ which enter the
normalized angular decay distribution. In fact, one obtains
($\hat H_{++}+\hat H_{00}+\hat H_{--}=1$)
\begin{eqnarray}\label{nlodecay}
\hat H_{++}&=&1+\frac{\alpha_s}{6\pi}\left(-5+(\pi^2+16)\sqrt{\mu_2}\right)
  +\ldots\nonumber \\
\hat H_{00}&=&0+\frac{\alpha_s}{6\pi}\left(4-2\pi^2\sqrt{\mu_2}\right)
  +\ldots\nonumber \\
\hat H_{--}&=&0+\frac{\alpha_s}{6\pi}\left(1+(\pi^2-16)\sqrt{\mu_2}\right)
  +\ldots
\end{eqnarray}
where we have used a small $\alpha_s$ expansion for the ratios
$H_{mm}/H_{U+L}$. For $W^+\to c\bar b$ the linear NLO quark mass effects now
amount to only $O(35\,\%)$ of the leading NLO contribution. The reason for the
reduction of the linear mass effects is that the largest linear mass effect
resides in the (unnormalized) polarized decay function $H_{++}$ which has a
sizeable LO contribution.

The normalized angular decay distribution~(\ref{Wtheta}) can be characterized
by the convexity parameter (see e.g.\ Ref.~\cite{Groote:2012xr})
\begin{equation}\label{convex}
c_f=\frac{d^2\widehat W(\theta)}{d(\cos\theta)^2}
  =\frac34(\hat\rho_{++}-2\hat\rho_{00}+\hat\rho_{--})
  (\hat H_{++}-2\hat H_{00}+\hat H_{--}).
\end{equation}
When $c_f$ is negative (positive), the angular decay distribution is described
by a downward (upward) open parabola. As a second global measure we introduce
the forward--backward asymmetry of the decay distribution defined by
\begin{equation}
A_{FB}=\frac{W(F)-W(B)}{W(F)+W(B)}
  =\frac34(\hat\rho_{++}-\hat\rho_{--})(\hat H_{++}-\hat H_{--}),
\end{equation}
where $W(F)=W(0\le\theta\le\pi/2)$ and $W(B)=W(\pi/2\le\theta\le\pi)$. If
there is an extremum of the angular decay distribution in the physical range
$-1\le\cos\theta\le 1$, the extremum is given by
\begin{equation}\label{extr}
\cos\theta\,\Big|_{\,\rm extr}=-\frac{A_{FB}}{c_f}
  =\,-\ \frac{(\hat\rho_{++}-\hat\rho_{--})}{(\hat\rho_{++}-2\hat\rho_{00}
+\hat\rho_{--})}
\ \frac{(\hat H_{++}-\hat H_{--})}{(\hat H_{++}-2\hat H_{00}+\hat H_{--})}.
\end{equation}
The three measures are not independent since
$\cos\theta\,\Big|_{\,\rm extr}=\,-A_{FB}/c_f$.

In the small $\alpha_s$ expansion and neglecting quark mass effects one has
\begin{eqnarray}
c_f&=&\frac34(1-3\hat\rho_{00})(1-12\,\frac{\alpha_s}{6\pi}),\\
A_{FB}&=&-\frac34(\hat\rho_{++}-\hat\rho_{--})(1-6\,\frac{\alpha_s}{6\pi}),\\
\cos\theta\,\Big|_{\,\rm extr}
&=&\ \frac{(\hat\rho_{++}-\hat\rho_{--})}{(1-3\hat\rho_{00})}
(1+6\,\frac{\alpha_s}{6\pi}).
\end{eqnarray}
The largest $\alpha_s$ correction occurs for the convexity parameter $c_f$.
Using $\alpha_s(m^2_W)=0.117$ one finds a $7.5\,\%$ reduction of $c_f$ through
the radiative corrections, i.e.\ the radiatively corrected angular decay
distribution becomes flatter by that amount. This flattening is clearly
discernible in the plot of the $\cos\theta$ distribution of the decay shown in
Sec.~8. 

Let us now for illustrative purposes turn to a specific example, namely the
cascade decay $t\to b+W^+(\to q_1\,\bar q_2)$. This process is particularly
interesting since the NLO radiative QCD corrections factorize into initial- and
final-state corrections, i.e.\ there is no NLO cross talk between top quark
decay and $W$ decay because of colour conservation~\cite{Brandenburg:2002xr}.

The spin density matrix elements of the $W^+$ in the decay process
$t\to b+W^+$ are well studied. At LO one has~\cite{Kane:1991bg}
\begin{eqnarray}\label{spinden}
\hat\rho_{++}({\it Born\/})&=&0\qquad\qquad\qquad\qquad\quad\to\,\,0.0007,
\nonumber\\[7pt]
\hat\rho_{00}({\it Born\/})&=&\frac1{1+2x^2}\ =0.696\qquad\,\to\,\,0.6887,
\nonumber\\[3pt]
\hat\rho_{--}({\it Born\/})&=&\frac{2x^2}{1+2x^2}\ =\ 0.304\qquad\!
  \to \,\,0.3106,
 \end{eqnarray}
where $x=m_W/m_t$. For the numerical values we use the central values of
$m_W=80.399\pm0.025\GeV$ and $m_t=172.0\pm0.9\pm1.3\GeV$ provided by the
Particle Data Group~\cite{Nakamura:2010zzi}. At leading order the density
matrix element $\hat\rho_{++}$ is not populated because of angular momentum
conservation in the two-body decay process. In Eq.~(\ref{spinden}) we have
also given the NLO QCD results indicated by arrows (cf.\
Refs.~\cite{Fischer:1998gsa,Fischer:2000kx,Fischer:2001gp,Do:2002ky}).%
\footnote{The NNLO corrections to the spin density matrix elements of the
$W^+$ have recently been calculated in Ref.~\cite{Czarnecki:2010gb}.} The
correction to $\hat\rho_{++}$ is very small. The absolute corrections to
$\hat\rho_{00}$ and $\hat\rho_{--}$ amount to $0.73\%$ and $0.66\%$ and are
thus considerably smaller than the final-state mass-zero corrections to
$\hat H_{++}$ and $\hat H_{00}$ given in Eq.~(\ref{nlodecay}). 

If a transverse direction can be specified, one can also probe the
nondiagonal spin density matrix elements $\hat\rho_{mm'}$ with $m\neq m'$.
The angular decay distribution is then given by~\cite{Korner:2003zq}
\begin{equation}
W(\theta)=\sum_{m,m',m''}\hat\rho_{mm'}\,d^1_{mm''}(\theta)\,
  d^1_{m'm''}(\theta)\,\, H_{m''m''}\,\,e^{-i(m-m')\phi},
\end{equation}
where $\phi$ denotes the azimuthal angle between the production and decay
plane. For $m'\ne m$ there will be the typical pattern of dispersive and
absorptive (or $CP$ violating) contributions proportional to
$\cos(m-m')\,\phi$ and $\sin(m-m')\,\phi$, respectively. We mention that, if
one generalizes the above example $t\to b+W^+(\to q_1\,\bar q_2)$ to the decay
of a polarized top quark $t(\uparrow)\to b+W^+(\to q_1\,\bar q_2)$, a
production plane can be defined with the help of the transverse polarization
of the top quark. The corresponding polar and azimuthal distributions
are given in Refs.~\cite{Fischer:1998gsa,Fischer:2001gp}. A further example
where the nondiagonal density matrix elements come into play is the much
discussed decay  $H\to f_{1}\bar f_{2}f_{3} \bar f_{4}$ treated e.g.\ in
Ref.~\cite{Bredenstein:2006rh,Bredenstein:2006ha} where one $f\bar f$ plane
provides the reference transverse direction needed for the definition of the
relative azimuthal orientation of the second plane.

\section{One-loop contributions}
For calculational reasons it is convenient to introduce linear combinations
of the diagonal helicity structure functions $H_{++}$, $H_{--}$ and $H_{00}$
given by
\begin{equation}\label{h123}
H_1=\frac12(H_{++}+H_{--}),\qquad H_2=\frac12(H_{++}-H_{--}),\qquad
H_3=\frac12(H_{++}+H_{--}-2H_{00}).
\end{equation}
The inverse relations read $H_{\pm\pm}=H_1\pm H_2$ and $H_{00}=H_1-H_3$.
Note that the linear combinations $H_2$ and $H_3$ appear as coefficients of
the $\cos\theta$ and $\cos^2\theta$ contributions in the angular decay
distribution~(\ref{Wtheta}).  

The one-loop QCD correction to the decay process $W^+\to q_1\,\bar q_2$ is
shown in Fig.~\ref{wlagu03}(b). The vertex correction to the Born-term
$(V-A)$ vertex factor
\begin{equation}
-i\frac{g_w}{\sqrt2}V_{ij}\gamma^\mu\frac{1-\gamma_5}2
\end{equation}
can be written as $-i(g_W/\sqrt2)V_{ij}\Delta\Gamma^\mu_L$. At NLO one finds
\begin{eqnarray}
\Gamma^\mu_L&=&\frac12\gamma^\mu(1-\gamma_5)+\Delta\Gamma_L^\mu
  \ =\ (1+A_L)\gamma^\mu\frac{1-\gamma_5}2
  +A_R\gamma^\mu\frac{1+\gamma_5}2\nonumber\\&&\strut
  +B_L^1p_1^\mu\frac{1-\gamma_5}2+B_R^1p_1^\mu\frac{1+\gamma_5}2
  +B_L^2p_2^\mu\frac{1-\gamma_5}2+B_R^2p_2^\mu\frac{1+\gamma_5}2
\end{eqnarray}
where, as in the LO case, $p_1$ and $p_2$ are the four-momentum of the up-type
quark and the down-type antiquark, respectively. The UV and IR singular parts
reside in the Born-term like structure $A_L$. In order to regularize the
singularities, we use dimensional regularization with $D=4-2\eps$. The UV
singularity is removed by UV renormalization while the IR singularity will be
cancelled by the corresponding contributions from the tree-graph contributions.
The form factors are in general complex valued, i.e.\ they contain absorptive
parts as can be visualized from Fig.~\ref{wlagu03}(b). For the present
calculation we only consider the diagonal helicity rate functions, and
thus we only need the real parts of the one-loop contributions. One has
\begin{eqnarray}
\real A_L&=&-\frac{\alpha_s}{4\pi}C_F\Gamma(1+\eps)
  \pfrac{4\pi\mu^2}{\sqrt{\mu_1\mu_2}q^2}^\eps
  \nonumber\\&&\kern-24pt\times
  \Bigg[\frac2\eps+2\frac{\mu_1+\mu_2-(\mu_1-\mu_2)^2}
  \sla\ln\pfrac{1-\tilde v}{1+\tilde v}
  +3\sla\ln\pfrac{1-\tilde v}{1+\tilde v}
  -(\mu_1-\mu_2)\ln\pfrac{\sqrt{\mu_1}}{\sqrt{\mu_2}}
  \nonumber\\&&\strut\kern-12pt
  +\frac2\sla(1-\mu_1-\mu_2)\left(\left(\frac1\eps-\ln\left(1-(\sqrt{\mu_1}
  -\sqrt{\mu_2})^2\right)\right)\ln\pfrac{1-\tilde v}{1+\tilde v}
  +\real L'\right)+4\Bigg],
  \nonumber\\
\real A_R&=&\frac{\alpha_s}{4\pi}C_F
  \Bigg[4\frac{\sqrt{\mu_1\mu_2}}\sla\ln\pfrac{1-\tilde v}{1+\tilde v}\Bigg],
  \nonumber\\
\real B_L^1&=&\frac{\alpha_s}{4\pi}C_F\frac{2m_1}{q^2}
  \Bigg[\frac{1-2\mu_1+(\mu_1-\mu_2)^2}\sla\ln\pfrac{1-\tilde v}{1+\tilde v}
  +(1-\mu_1+\mu_2)\ln\pfrac{\sqrt{\mu_1}}{\sqrt{\mu_2}}+1\Bigg],\nonumber\\
\real B_R^1&=&\!\!\!-\frac{\alpha_s}{4\pi}C_F\frac{2m_2}{q^2}
  \nonumber\\&&\times
  \Bigg[\frac{1-\mu_1-\mu_2+(1+\mu_1-\mu_2)^2}
  \sla\ln\pfrac{1-\tilde v}{1+\tilde v}
  -(2+\mu_1-\mu_2)\ln\pfrac{\sqrt{\mu_1}}{\sqrt{\mu_2}}+1\Bigg],\nonumber\\
\real B_L^2&=&\frac{\alpha_s}{4\pi}C_F\frac{2m_1}{q^2}\nonumber\\&&\times
  \Bigg[\frac{1-\mu_1-\mu_2+(1-\mu_1+\mu_2)^2}
  \sla\ln\pfrac{1-\tilde v}{1+\tilde v}
  -(2-\mu_1+\mu_2)\ln\pfrac{\sqrt{\mu_2}}{\sqrt{\mu_1}}+1\Bigg],\nonumber\\
\real B_R^2&=&\!\!\!-\frac{\alpha_s}{4\pi}C_F\frac{2m_2}{q^2}
  \nonumber\\&&\times
  \Bigg[\frac{1-2\mu_2+(\mu_1-\mu_2)^2}\sla\ln\pfrac{1-\tilde v}{1+\tilde v}
  +(1+\mu_1-\mu_2)\ln\pfrac{\sqrt{\mu_2}}{\sqrt{\mu_1}}+1\Bigg]\qquad
\end{eqnarray}
where $C_F=(N_c^2-1)/2N_c=4/3$. We have introduced a velocity parameter
$\tilde v$ defined by
\begin{equation}
\label{velo}
\tilde v=\sqrt{\frac{1-(\sqrt{\mu_1}+\sqrt{\mu_2}\,)^2}{1
  -(\sqrt{\mu_1}-\sqrt{\mu_2}\,)^2}}\,\,.
\end{equation}
{\it Per se\/} the velocity parameter has no physical meaning except that it
reduces to the usual velocity $v=\sqrt{1-4m^2/q^2}$ in the equal mass limit.
The function $\real L'$ is given in Appendix~A. The scale $\mu$ in $\real A_L$
has been introduced to keep the strong coupling constant dimensionless in
$D=4-2\eps$ dimensions. The dependence on $\mu$ cancels in the sum of the
one-loop and tree-graph contributions. The one-loop contributions to the
helicity structure functions finally read
\begin{eqnarray}
H_1({\it loop\/})&=&8N_cq^2(1-\mu_1-\mu_2)\real A_L
  +16N_cq^2\sqrt{\mu_1\mu_2}\real A_R,\nonumber\\[7pt]
H_2({\it loop\/})&=&-8N_cq^2\sla\real A_L,\nonumber\\[7pt]
H_3({\it loop\/})&=&8N_cq^2\lambda\real A_L\nonumber\\[7pt]&&
  +4N_cq^2\lambda\left(m_1(\real B_L^1-\real B_L^2)
  +m_2(\real B_R^1-\real B_R^2)\right).\qquad
\end{eqnarray}

\begin{figure}
\begin{center}
\epsfig{figure=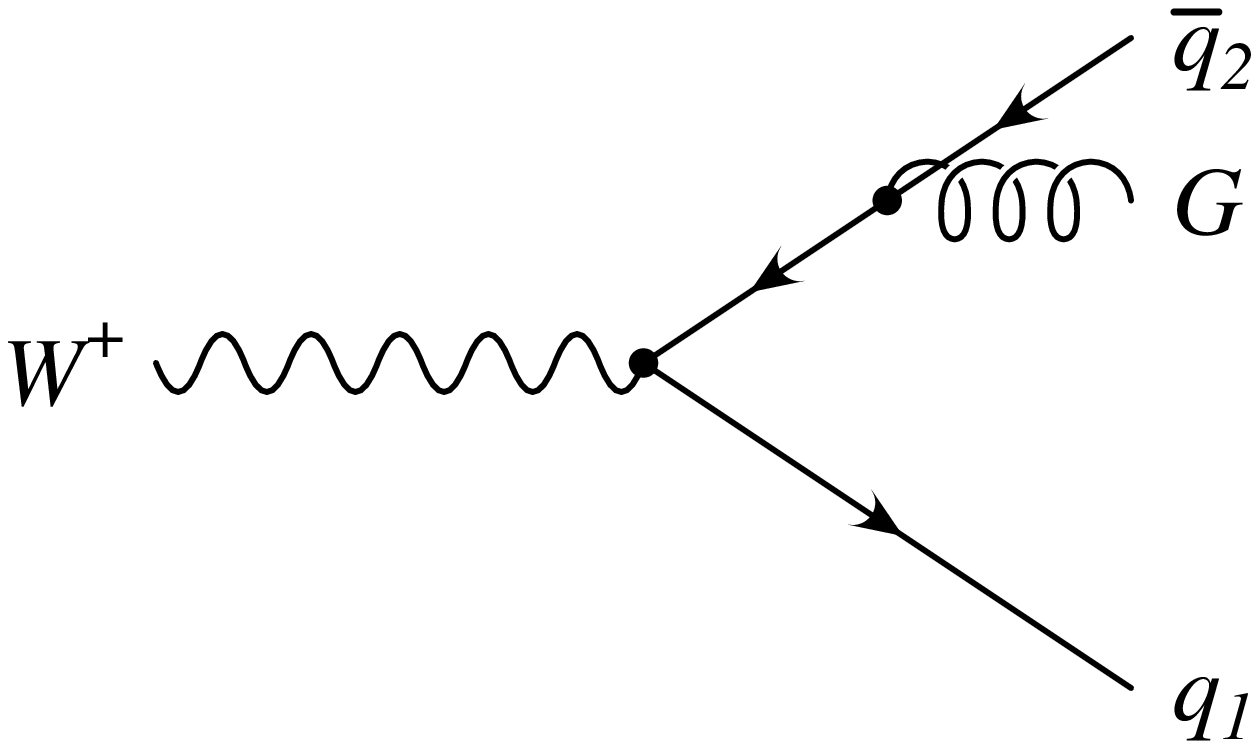, scale=0.5}\qquad
\epsfig{figure=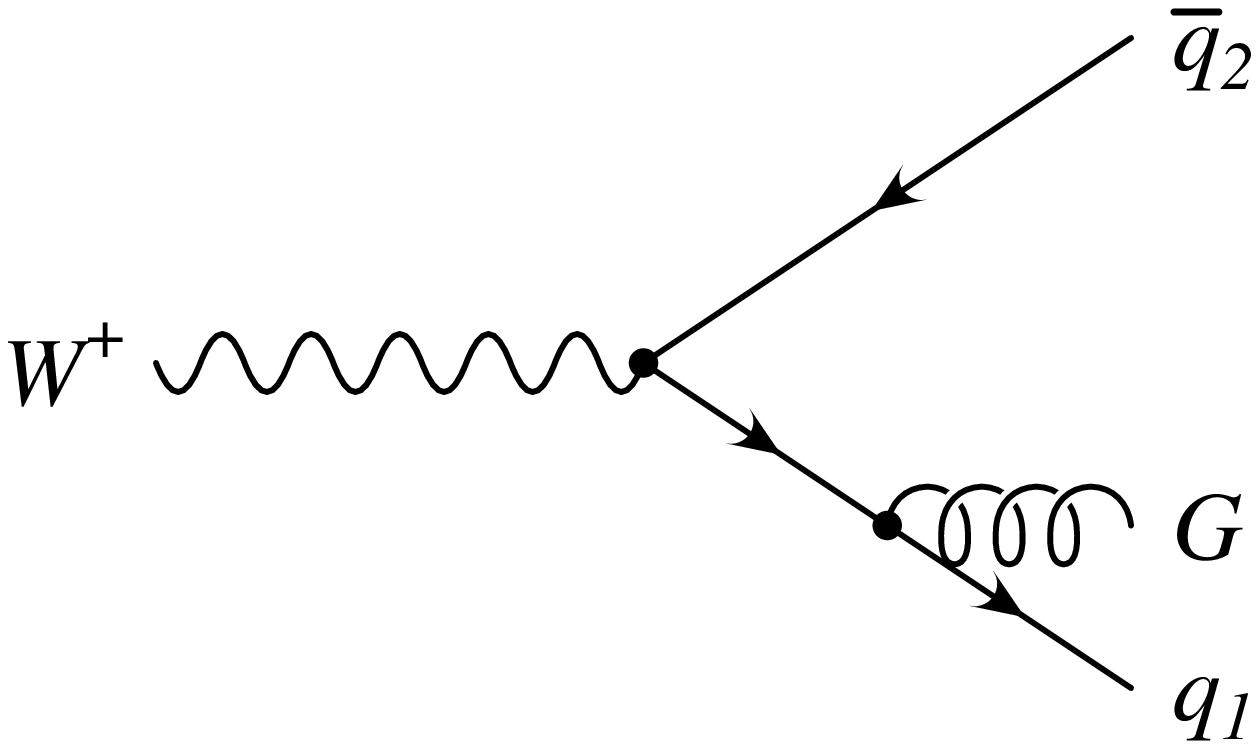, scale=0.5}
\end{center}
\caption{\label{wlagu21}Feynman diagrams for the NLO tree-graph contributions
to the decay process $W^+\to q_1 \bar{q}_2$}
\end{figure}

\section{Tree-graph contributions}
In accordance with the Lee--Nauenberg theorem, the IR singularities of the
one-loop contribution will have to cancel against the gluon-emission
tree-graph contributions depicted in Fig.~\ref{wlagu21}. The result of the
full phase-space integration can be expressed in terms of the decay rate terms
$\ell_0,\ldots,\ell_4$, $I_z^\ell(0)$, $S_z^\ell(0)$, $I_1^\ell(0)$,
$S_1^\ell(0)$, and $I^\ell(0)$ listed in Appendix~A. Again we list our results
in terms of the auxiliary expressions $H_1$, $H_2$ and $H_3$ defined in
Eq.~(\ref{h123}). One has
\begin{eqnarray}
H_1({\it tree\/})&=&N\Big[4(1-\mu_1-\mu_2)D_S
  -4\mu_1(1+7\mu_1-\mu_2)I^\ell_1(0)\strut\nonumber\\[3pt]&&\strut
  -2\sqrt{\mu_1}(1-12\mu_1-2\mu_2-5\mu_1^2+4\mu_1\mu_2+\mu_2^2)S^\ell_1(0)
  \strut\nonumber\\[6pt]&&\strut
  -2\mu_1(6+4\mu_1-7\mu_2)\ell_1+2\mu_2(2+3\mu_1)\ell_2-2(1-11\mu_1+\mu_2)\sla
  \Big],\nonumber\\[7pt]
H_2({\it tree\/})&=&N\Big[-4\sla D_I
  +4(1-3\mu_1-2\mu_2-\mu_1^2+\mu_2^2)I^\ell(0)\strut\nonumber\\[3pt]&&\strut
  -2(2-\mu_1+\mu_2-\mu_1^2+\mu_1\mu_2)\ell_0-8\lambda\ell_4
  \strut\nonumber\\[7pt]&&\strut
  +4\sla(1+2\mu_1-\mu_2)\ell_1+2\sla(2+\mu_1+\mu_2)\ell_2
  \strut\nonumber\\[6pt]&&\strut
  +\left(3+14\sqrt{\mu_1}-3\mu_1+3\mu_2\right)
  \left((1-\sqrt{\mu_1})^2-\mu_2\right)\Big],\nonumber\\[7pt]
H_3({\it tree\/})&=&N\Big[4\lambda D_S
  -12\mu_1(1+7\mu_1-\mu_2)I^\ell_1(0)\strut\nonumber\\[3pt]&&\strut
  -6\sqrt{\mu_1}(1-12\mu_1-2\mu_2-5\mu_1^2+4\mu_1\mu_2+\mu_2^2)S^\ell_1(0)
  \strut\nonumber\\[7pt]&&\strut
  -2\mu_1(20+13\mu_1-24\mu_2+\mu_1^2+\mu_1\mu_2+4\mu_2^2)\ell_1
  \strut\nonumber\\[7pt]&&\strut
  +2\mu_2(4+12\mu_1-\mu_2-4\mu_1^2-\mu_1\mu_2-\mu_2^2)\ell_2
  \strut\nonumber\\[6pt]&&\strut
  -2(3-36\mu_1-\mu_1^2+8\mu_1\mu_2-\mu_2^2)\sla\Big],
\end{eqnarray}
where 
\begin{equation}\label{defN}
N:=\alpha_sN_cC_Fq^2/(\pi\sla),
\end{equation}
\begin{eqnarray}
D_S&:=&(1-\mu_1-\mu_2)\left(D^\ell+S^\ell_z(0)\right)-2\sla D\nonumber\\[3pt]&&
  +\frac34\left((1+\mu_1-\mu_2)\ell_1+(1-\mu_1+\mu_2)\ell_2+\sla\right),
  \nonumber\\
D_I&:=&(1-\mu_1-\mu_2)\left(D^\ell+I^\ell_z(0)\right)-2\sla D\nonumber\\[3pt]&&
  +\frac34\left((1+\mu_1-\mu_2)\ell_1+(1-\mu_1+\mu_2)\ell_2+\sla\right).\qquad
\end{eqnarray}
We have isolated the IR singular parts in $D$ and $D^\ell$ given by
\begin{eqnarray}
D&:=&\ln\pfrac{\lambda}{\sqrt{\Lambda\mu_1\mu_2}}-1,\nonumber\\
D^\ell&:=&\ln\pfrac{\lambda}{\sqrt{\Lambda\mu_1\mu_2}}\ln\alpha_+
  +\frac12\Li_2(1-\alpha_+)-\frac12\Li_2(1-\alpha_-)
\end{eqnarray}
with $\alpha_+=(1-\mu_1-\mu_2+\sla)/(1-\mu_1-\mu_2-\sla)=\alpha_-^{-1}$. The
IR singularity has been regularized by a small but finite gluon mass
$m_G=\sqrt{\Lambda q^2}$. Since the one-loop calculation has been done using
dimensional regularization, one needs to convert the IR divergent piece of
the tree-graph contribution to the corresponding expression in dimensional
regularization by using the one-loop relation
\begin{equation}
\ln\Lambda=\pfrac{\mu^2}{q^2}^\eps\left(\frac1\eps-\gamma_E+\ln(4\pi)\right).
\end{equation}

\section{Total NLO contribution}
Because of the aforementioned Lee--Nauenberg theorem, the IR singularities
cancel when adding the one-loop and tree-graph contributions. Using the IR
finite quantities
\begin{eqnarray}\label{ASI}
A_S&:=&D_S+\frac{q^2}{2N}\real A_L\nonumber\\
  &=&\frac12(1-\mu_1-\mu_2)\left(t_A+2S^\ell_z(0)\right)-\sla\ell_A
  +\frac12\left(1-\mu_1-\mu_2+\frac12\lambda\right)\ell_3
  \strut\nonumber\\&&\strut
  +\frac14(\mu_1-\mu_2)\sla\ell_B
  +\frac34\left((1+\mu_1-\mu_2)\ell_1+(1-\mu_1+\mu_2)\ell_2+\sla\right),
  \nonumber\\[7pt]
A_I&:=&D_I+\frac{q^2}{2N}\real A_L\nonumber\\
  &=&\frac12(1-\mu_1-\mu_2)\left(t_A+2I^\ell_z(0)\right)-\sla\ell_A
  +\frac12\left(1-\mu_1-\mu_2+\frac12\lambda\right)\ell_3
  \strut\nonumber\\&&\strut
  +\frac14(\mu_1-\mu_2)\sla\ell_B
  +\frac34\left((1+\mu_1-\mu_2)\ell_1+(1-\mu_1+\mu_2)\ell_2+\sla\right)
\end{eqnarray}
($\ell_A$, $\ell_B$ and $t_A$ are listed in Appendix~A), the total results
read
\begin{eqnarray}
H_1(\alpha_s)&=&N\Big[4(1-\mu_1-\mu_2)A_S-4\mu_1(1+7\mu_1-\mu_2)I^\ell_1(0)
  \strut\nonumber\\[6pt]&&\strut
  -2\sqrt{\mu_1}(1-12\mu_1-5\mu_1^2-2\mu_2+4\mu_1\mu_2+\mu_2^2)S^\ell_1(0)
  \strut\nonumber\\[7pt]&&\strut
  -2\mu_1(6+4\mu_1-7\mu_2)\ell_1+2\mu_2(2+3\mu_1)\ell_2
  \strut\nonumber\\[6pt]&&\strut
  -8\mu_1\mu_2\ell_3-2(1-11\mu_1+\mu_2)\sla\Big],\\[7pt]
H_2(\alpha_s)&=&N\Big[-4\sla A_I
  +4(1-3\mu_1-\mu_1^2-2\mu_2+\mu_2^2)I^\ell(0)
  \strut\nonumber\\[6pt]&&\strut
  -2(2-\mu_1-\mu_1^2+\mu_2+\mu_1\mu_2)\ell_0-8\lambda\ell_4
  \strut\nonumber\\[7pt]&&\strut
  +4\sla(1+2\mu_1-\mu_2)\ell_1+2\sla(2+\mu_1+\mu_2)\ell_2
  \strut\nonumber\\[6pt]&&\strut
  +\left(3+14\sqrt{\mu_1}-3\mu_1+3\mu_2\right)
  \left((1-\sqrt{\mu_1})^2-\mu_2\right)\Big],\\[7pt]
H_3(\alpha_s)&=&N\Big[4\lambda A_S-12\mu_1(1+7\mu_1-\mu_2)I^\ell_1(0)
  \strut\nonumber\\[6pt]&&\strut
  -6\sqrt{\mu_1}(1-12\mu_1-5\mu_1^2-2\mu_2+4\mu_1\mu_2+\mu_2^2)S^\ell_1(0)
  \strut\nonumber\\[7pt]&&\strut
  -2\mu_1(20+13\mu_1+\mu_1^2-24\mu_2+\mu_1\mu_2+4\mu_2^2)\ell_1
  \strut\nonumber\\[7pt]&&\strut
  +2\mu_2(4+12\mu_1-4\mu_1^2-\mu_2-\mu_1\mu_2-\mu_2^2)\ell_2
  \strut\nonumber\\[4pt]&&\strut
  +\lambda\left(\mu_1+\mu_2-(\mu_1-\mu_2)^2\right)\ell_3
  -(\mu_1-\mu_2)\lambda\sla\ell_B
  \strut\nonumber\\[6pt]&&\strut
  -2(3-36\mu_1-\mu_1^2+8\mu_1\mu_2-\mu_2^2)\sla\Big].
\end{eqnarray}
For $\mu_1=\mu_2$ we agree with our previous NLO QCD results on
$(\gamma^*,Z)(\uparrow)\to q\bar q$~\cite{Groote:1995yc,Groote:1995ky,%
Groote:1996nc,Groote:2008ux,Groote:2009zk}.

A further check can be done by comparing the sum of the partial helicity
structure functions $H_{U+L}=H_{++}+H_{00}+H_{--}=3H_1-H_3$ with the
corresponding results~\cite{Denner:1990tx,Denner:1991kt}. For the
unpolarized decay function $H_{U+L}(\alpha_s)$ we obtain
\begin{eqnarray}\label{upluslals}
H_{U+L}(\alpha_s)&=&N\Big[4(3(1-\mu_1-\mu_2)-\lambda)A_S
  \strut\nonumber\\[6pt]&&\strut
  +2\mu_1\left(2+\mu_1+\mu_1^2-18\mu_2+\mu_1\mu_2+4\mu_2^2\right)\ell_1
  \strut\nonumber\\[8pt]&&\strut
  +2\mu_2\left(2-18\mu_1+4\mu_1^2+\mu_2+\mu_1\mu_2+\mu_2^2\right)\ell_2
  \strut\nonumber\\[8pt]&&\strut
  -\left((1-\mu_1-\mu_2-\lambda)\lambda-6\mu_1\mu_2\right)
  \ell_3+(\mu_1-\mu_2)\lambda\sla\ell_B
  \strut\nonumber\\[6pt]&&\strut
  +2\left(1-5\mu_1-5\mu_2-\lambda+6\mu_1\mu_2\right)\sla\Big]
\end{eqnarray}
in full agreement with Ref.~\cite{Denner:1991kt}.\footnote{We also find
agreement with the final result in Ref.~\cite{Denner:1990tx} after correcting
two typos in Eq.~(A.50) of Ref.~\cite{Denner:1990tx}, namely after removing
the denominator factors $(1+w_1)$ in two of the Spence functions in~(A.50). We
thank A.~Denner for a communication on these typographical errors.}

\section{High-energy and threshold limit}
Since our results are obtained in analytical form, one can study different 
limiting cases. In the high-energy (or mass-zero) limit one needs to
expand the K\"all\'en function up to $O(\mu_i^2)$. One has
\begin{equation}\label{helimit}
\sqrt\lambda=\sqrt{1+\mu_1^2+\mu_2^2-2\mu_1-2\mu_2-2\mu_1\mu_2}
  \ =\ 1-\mu_1-\mu_2-\mu_1\mu_2+O(\mu_i^3).
\end{equation}
The high-energy limit of the decay rate terms are given in Appendix~B. One has
\begin{eqnarray}
H_{++}(\alpha_s)&=&H_1(\alpha_s)-H_2(\alpha_s)
  \ \to\ 8N_cq^2\left\{\frac{\alpha_s}{6\pi}\right\},\nonumber\\
H_{00}(\alpha_s)&=&H_1(\alpha_s)-H_3(\alpha_s)
  \ \to\ 8N_cq^2\left\{\frac{4\alpha_s}{6\pi}\right\},\nonumber\\
H_{--}(\alpha_s)&=&H_1(\alpha_s)+H_2(\alpha_s)
  \ \to\ 8N_cq^2\left\{1+\frac{\alpha_s}{6\pi}\right\}.
\end{eqnarray}
This result has already been used in Sec.~3.

At threshold one has $\sqrt{\mu_1}+\sqrt{\mu_2}\to 1$ and thus $\lambda\to 0$.
Using the results of Appendix~C one obtains up to $O(\alpha_s)$
\begin{equation}\label{hthresh}
H_{++}=H_{00}=H_{--}=H_{tt}\to 8N_cq^2\left\{\sqrt{\mu_1\mu_2}
  +8\pi^2\frac{\alpha_s}{3\pi\sla}\mu_1\mu_2\right\}.
\end{equation}
At threshold, all four $O(\alpha_s)$ helicity rate functions are equal
to one another as is true at LO (see the pertinent discussion in Sec.~2).
Concerning the on-shell decay of the $W^+$ involving the polarized decay
functions $H_{++}=H_{00}=H_{--}$ one thus has a flat angular decay
distribution at threshold also at NLO. The Coulomb singularity proportional to
$1/\sla$ in Eq.~(\ref{hthresh}) signals that perturbation theory breaks down
close to threshold. One has to use nonperturbative methods to analyze the
region close to threshold similar to the analysis of
$e^+e^-\to\gamma,Z\to t\bar t$ close to threshold discussed in
Refs.~\cite{Fadin:1987wz,Harlander:1996vg,Awramik:2001ut}.

\section{Numerical results for off-shell and on-shell\\
  polarized decay functions}
In this section we present our numerical NLO results for the three helicity
rate functions $H_{mm}$ for on-shell and off-shell $W$ bosons. We choose the
$\sqrt{q^2}$ range to extend from threshold $\sqrt{q^2}=m_b+m_c$ to the
maximal energy $\sqrt{q^2}=m_t-m_b$ attainable in the decay $t\to b+W^+$. In
order to highlight quark mass effects we take the decay channel with the
highest quark masses, namely the channel $W^+\to c\bar b$ proportional to
$(V_{cb})^2=(0.041)^2$. For the quark masses we take the pole masses  
$m_t=172.0\GeV$, $m_b=4.8\GeV$ and $m_c=1.5\GeV$. We let $\alpha_s$ run
with two-loop accuracy. At $q^2=m_W^2=80.385\GeV^2$ we have $\alpha_s=0.117$. 

\begin{figure}
\begin{center}
\epsfig{figure=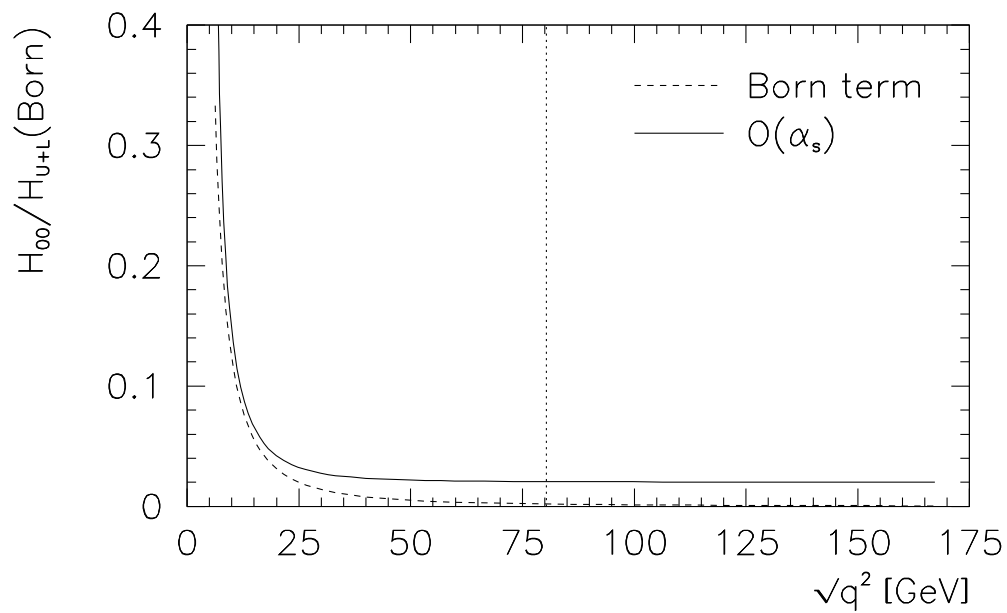, scale=0.8}
\end{center}
\caption{\label{ener00f}Energy dependence of the normalized coefficient
  $H_{00}/H_{U+L}({\it Born\/})$ for the $(c \bar b)$ case in the interval 
$[m_b+m_c,m_t-m_b]$ at
  LO (dashed lines) and NLO (solid lines). The dotted vertical line in
  Figs.~\ref{ener00f}--\ref{relplotf} marks the position of 
  an on-shell $W$ boson.}
\end{figure}

\begin{figure}
\begin{center}
\epsfig{figure=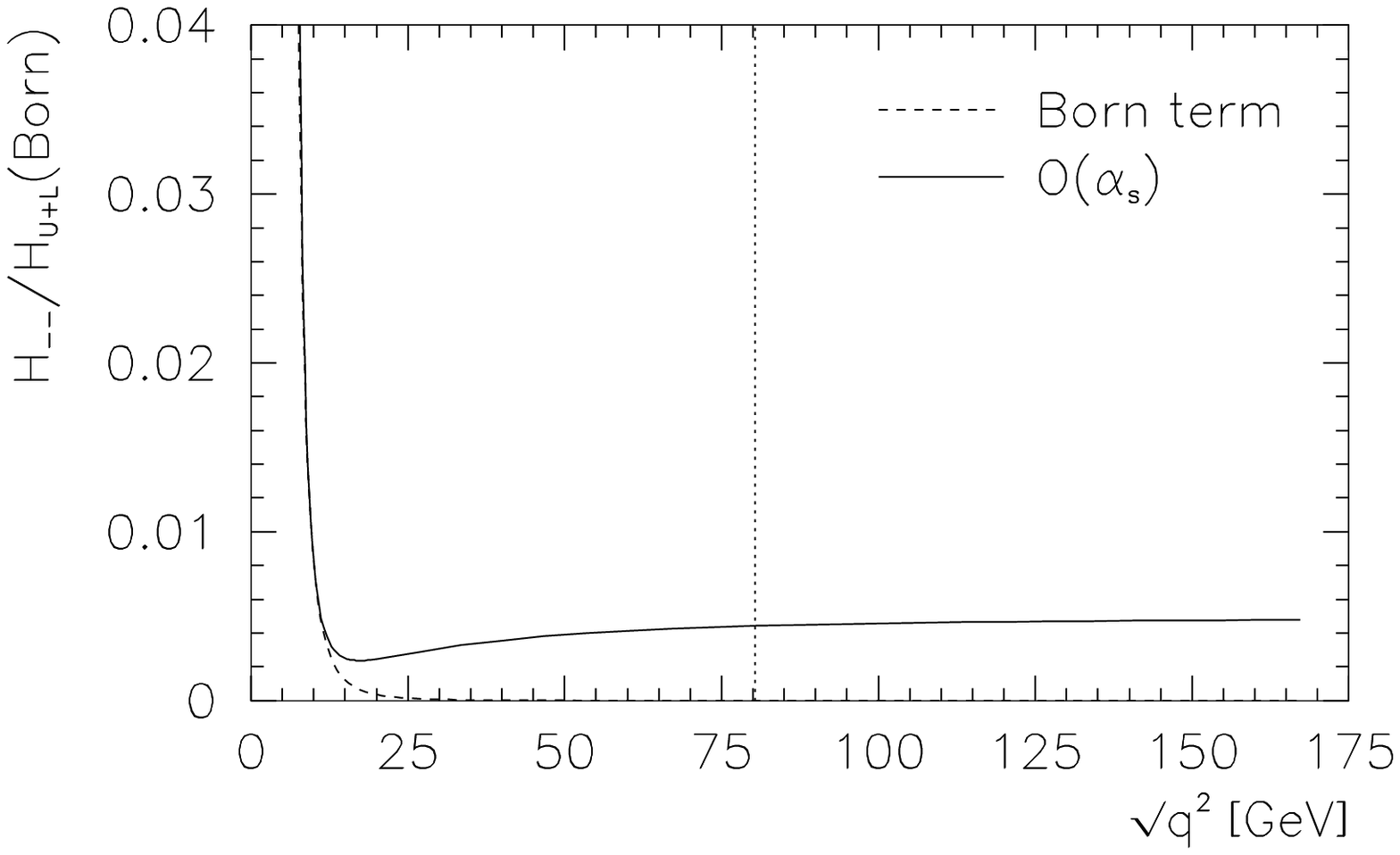, scale=0.8}
\end{center}
\caption{\label{enermmf}Energy dependence of the normalized coefficient
  $H_{--}/H_{U+L}({\it Born\/})$ for the $(c \bar b)$ case in the interval 
$[m_b+m_c,m_t-m_b]$ at
  LO (dashed lines) and NLO (solid lines)}
\end{figure}

\begin{figure}
\begin{center}
\epsfig{figure=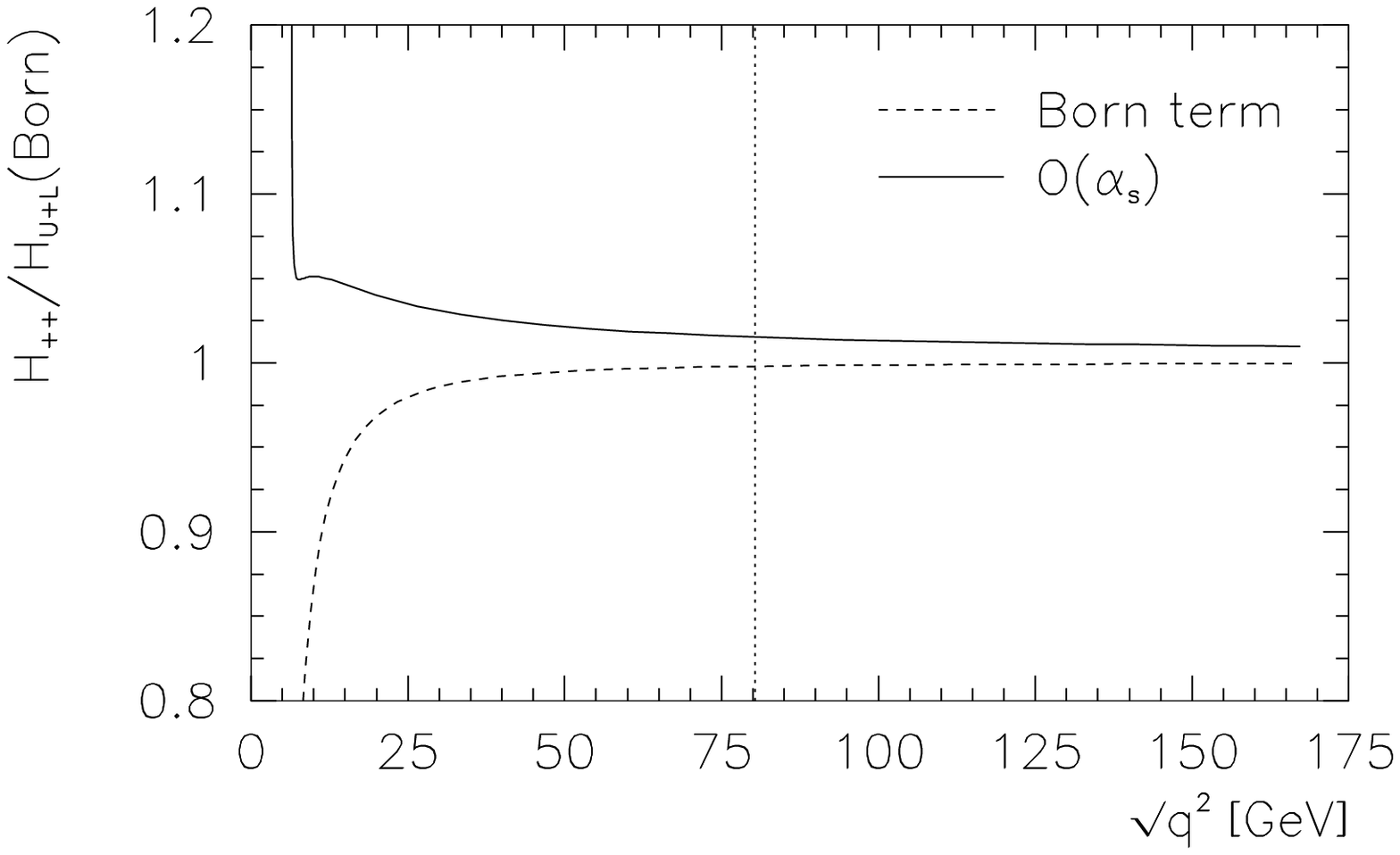, scale=0.8}
\end{center}
\caption{\label{enerppf}Energy dependence of the normalized coefficient
  $H_{++}/H_{U+L}({\it Born\/})$ for the $(c \bar b)$ case in the interval 
$[m_b+m_c,m_t-m_b]$ at
  LO (dashed lines) and NLO (solid lines)}
\end{figure}

In Figs.~\ref{ener00f}, \ref{enermmf} and~\ref{enerppf} we display the
$\sqrt{q^2}$ dependence of the Born-term and $O(\alpha_s)$ helicity rate 
functions $H_{00}$, $H_{--}$ and $H_{++}$ for the process $W^+\to c\bar b$.
We choose to normalize our results to the unpolarized Born-term rate function
$H_{U+L}({\it Born\/})$ given in Eq.~(\ref{hupBorn}).

Fig.~\ref{ener00f} shows that the ratio
$H_{00}({\it Born\/})/H_{U+L}({\it Born\/})$ rapidly approaches the
appropriate threshold value of $1/3$ at the lower end of the spectrum. The
corresponding NLO ratio quickly approaches $+\infty$ at threshold because of
the Coulomb singularity in the $\alpha_s$ NLO one-loop contribution. Towards
the higher end of the $\sqrt{q^2}$ spectrum the two ratios quickly reach their
respective asymptotic values of zero and $2\alpha_s/3\pi$. For the maximal
energy $\sqrt{q^2}=m_t-m_b$ the results are already close to the high-energy
limit. The Born-term result approaches zero while the $O(\alpha_s)$
result stays at a finite value $2\alpha_s/3\pi\approx 0.02$
(with $\alpha_s(m_t-m_b)\approx 0.1$). For $H_{--}({\it Born\/})$
Fig.~\ref{enermmf} shows that, at maximal energy, the high-energy result
$\alpha_s/6\pi\approx 0.005$ is already obtained with high accuracy while the
Born-term result again approaches zero. Finally, for the normalized
coefficient $H_{++}({\it Born\/})$ one sees from Fig.~\ref{enerppf} that the
Born-term result approaches the value $1$ at maximal energy. 

All three plots show that the approach to the high-energy (or mass-zero) limit
is rather slow for the $\alpha_s$ corrections. In particular one is not close
to the asymptotic NLO values
$H_{--}/H_{U+L}({\it Born\/})\sim(1+\alpha_s/6\pi)$,
$H_{00}/H_{U+L}({\it Born\/})\sim 4\alpha_s/6\pi$ and 
$H_{++}/H_{U+L}({\it Born\/})\sim\alpha_s/6\pi$ at the on-shell value
$\sqrt{q^2}=m_W$ indicated by the dotted vertical lines in
Figs.~\ref{ener00f}, \ref{enermmf} and~\ref{enerppf}. The large NLO mass
effects even at the scale $\sqrt{q^2}=m_W$ have been discussed before in
Sec.~3 and in Ref.~\cite{Groote:2012xr} where one can find an $O(\mu_i)$
expansion of the NLO mass effects.

\begin{figure}
\begin{center}
\epsfig{figure=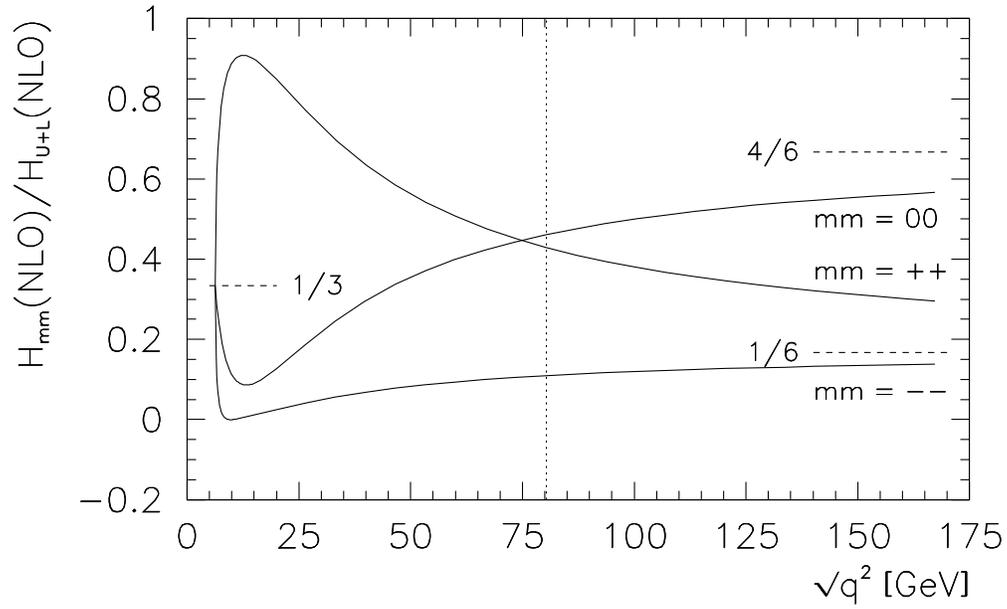, scale=0.8}
\end{center}
\caption{\label{relplotf}Energy dependence of the NLO corrections to
  $H_{mm}({\rm NLO})$ ($m=\pm,0$), divided by
  $H_{U+L}({\rm NLO})=H_{--}({\rm NLO})+H_{00}({\rm NLO})+H_{++}({\rm NLO})$
  for the $(c \bar b)$ case in the interval $[m_b+m_c,m_t-m_b]$}
\end{figure}

In Fig.~\ref{relplotf} we leave out the Born-term contributions and show the
NLO corrections to $H_{mm}({\rm NLO})$, divided by the sum of these. It is
obvious that, at threshold, the effect of the Coulomb singularity drops out in
this ratio and all three helicity structure functions contribute with a
relative factor $1/3$. On the other end of the spectrum in Fig.~\ref{relplotf}
the curves start their slow approach to the limiting values $1/6$ (for
$H_{\pm\pm}$) and $4/6$ (for $H_{00}$).

\begin{table}[ht]\begin{center}
\begin{tabular}{|l||c|c|c|c|}\hline
&Born\ $m_{i}=0$ &Born\ $m_{i}\neq 0$
&$O(\alpha_{s})\ m_{i}=0$&$O(\alpha_{s})\ m_{i}\neq 0$\\\hline
$W^{+}\to c\bar b$&&&&\\
$c_f$&$-0.8142$&$-0.8095$&$-0.7348$&$-0.7466$\\[-2.0ex]
$A_{FB}$&$-0.2280$&$-0.2276$&$-0.2234$&$-0.2253$\\[-2.0ex]
$\cos\theta\,|_{\,\rm max}$&$-0.2800$&$-0.2811$&$-0.3040$&$-0.3018$\\\hline
$W^{+}\to c\bar s$&&&&\\
$c_f$&$-0.8142$&$-0.8138$&$-0.7348$&$-0.7352$\\[-2.0ex]
$A_{FB}$&$-0.2280$&$-0.2280$&$-0.2234$&$-0.2235$\\[-2.0ex]
$\cos\theta\,|_{\,\rm max}$&$-0.2800$&$-0.2801$&$-0.3040$&$-0.3039$\\\hline
\end{tabular}
\caption{\label{tab1}The measures $c_f$, $A_{FB}$ and $\cos\theta\,|_{\rm\,max}$
for LO and NLO results at $q^2=m_W^2$ for the cascade process
$t\to b+W^+(\to c\bar b,\ c\bar s)$. Shown are massless results as well as
results where the quark masses ($m_s=150\GeV$, $m_c=1.5\GeV$ and $m_b=4.8\GeV$)
are taken into account.}
\end{center}\end{table}

\begin{figure}\begin{center}
\epsfig{figure=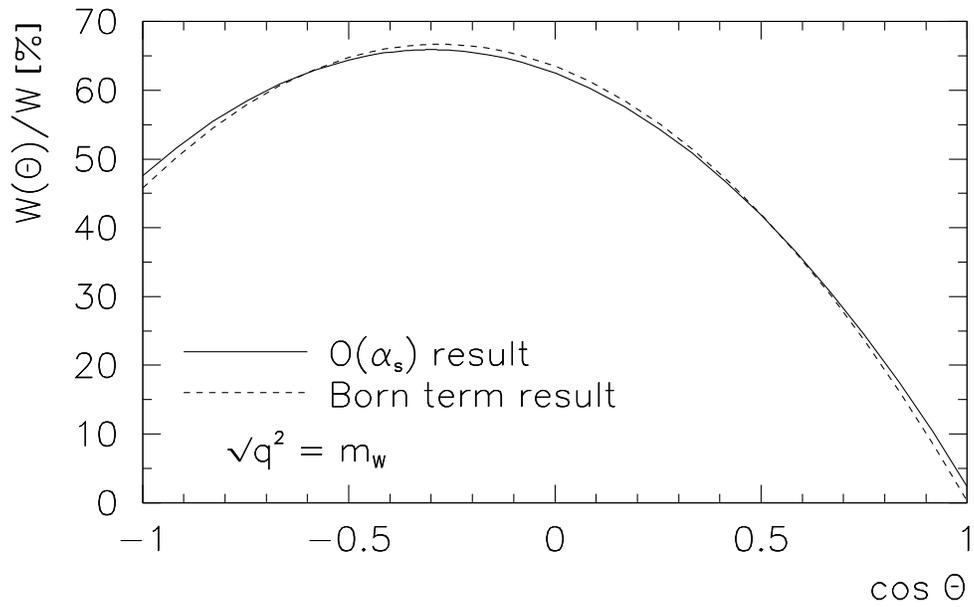, scale=0.8}
\caption{\label{nurks}Normalized angular decay distribution
$\widehat W(\theta)=W(\theta)/W$ at LO (dashed line) and NLO (full line) for
the on-shell decay $t\to b+W^{+}(\to c\bar b)$. The NLO result contains both
initial-state and final-state corrections}
\end{center}\end{figure}

In Fig.~\ref{nurks} we plot the $\cos\theta$ distribution for
$\widehat W(\theta)$. It is quite apparent that the distribution becomes
flatter through the radiative corrections. Numerical values for the parameters
$c_f$, $A_{FB}$ and $\cos\theta\,|_{\,\rm max}$ can be found in
Tab.~\ref{tab1}. The negative value of the convexity parameter $c_f$ means
that the angular decay distribution is given by a downward-open parabola.
Quark mass effects can be seen to be almost negligibly small for the
$W^+\to c\bar s$ channel.

We assume that it would be experimentally feasible to flavour-tag bottom and
charm quark jets, at least for a large fraction of the corresponding top quark
decays. If the hadronic flavour channel cannot be isolated, one has to take
the appropriate flavour sums using the unitarity of the Kobayashi--Maskawa
matrix. Furthermore, in the untagged case, the parity violating contribution
proportional to $\cos\theta$ would drop out and the angular decay distribution
would become symmetric in $\cos\theta$. The resulting polar decay distribution
reads
\begin{equation}
\widehat W_{ut}(\cos\theta)=\frac12\left(\,\widehat W_{t}(\cos\theta)+ 
\widehat W_{t}(-\cos\theta)\,\right)
\end{equation}
where ``$t$'' and ``$ut$'' stand for ``flavour tagged'' and ``flavour
untagged''.

Finite $W$-width effects in top quark decays have been considered in
Refs.~\cite{Jezabek:1988iv,Jezabek:1993wk,Do:2002ky} (see also
Ref.~\cite{Denner:2012yc}). We have recalculated the finite width correction
to the total top quark width using the mass values of the present paper 
and find that the total width is reduced by $1.55\,\%$ by the finite width
corrections. We also found that the longitudinal and transverse widths are
reduced by $1.35\,\%$ and $1.99\,\%$, resp., similar to the corresponding
values found in Ref.~\cite{Do:2002ky}. Curiously enough, the respective finite
width corrections are almost completely cancelled by the positive
contributions of the perturbative electroweak
corrections~\cite{Jezabek:1988iv,Jezabek:1993wk,Do:2002ky} such that these
corrections taken together will affect the angular decay distributions
only in a minor way.

\section{The decays $H\to W^-+W^{\ast+}(\to q_1\bar q_2)$\\
and $H\to Z+Z^{\ast}(\to q\bar q)$}
In this section we consider quark mass and off-shell effects in the polar
angle distribution of the decay $W^{\ast+}(\uparrow)\to q_1\bar q_2$ where the
off-shell $W^{\ast+}$ is produced in the Higgs decay $H\to W^-+W^{\ast+}$. We
shall also briefly touch on the subject of the three-body decay
$H\to Z+Z^\ast(\to q\bar q)$. The corresponding leptonic modes have recently
been observed at the LHC and are therefore adequately dubbed ``Higgs discovery
channels''~\cite{:2012gk,:2012gu}. Off-shell effects in these decays will 
lead to additional scalar and scalar--longitudinal interference contributions
in e.g.\ the off-shell decay $W^{\ast+}(\uparrow)\to q_1\bar q_2$ well
familiar from neutron beta decay and from the semileptonic decay
$\Xi^0\to\Sigma^++\mu^-\bar\nu_\mu$~\cite{Kadeer:2005aq}, or from the decay
$B\to D^{(\ast)}+\tau^-\bar\nu_\tau$~\cite{Korner:1989qb}. The scalar and
scalar--longitudinal interference contributions are quadratic in the quark
masses and can thus be neglected at the scale $q^2=m_W^2$. However, for the
off-shell decay $H\to W^-+W^{\ast+}$ the scale is not set by $m_W^2$ but by
the off-shellness of the $W^{\ast+}$ which extends from threshold
$q^2=(m_1+m_2)^2$ (maximal recoil point) to the zero recoil point at
$q^2=(m_H-m_W)^2$, i.e.\ one has ($m_H=126\GeV$)
\begin{equation}
(m_1+m_2)^2\le q^2\le(m_H-m_W)^2.
\end{equation}
One will therefore have to carefully consider quark mass and $W^{\ast+}$ 
off-shell effects in the $q^2$ region close to threshold.

The differential decay distribution for the decay
$H\to W^-W^{\ast\,+}(\to q_1\bar q_2)$ is given by
\begin{equation}
\label{hdecay}
\frac{d\Gamma}{dq^2d\cos\theta}=\frac{g_{w}^4}{1024\pi^3}|V_{12}|^2
  \frac{|\vec{p}_W||\vec{p}|}{m_{H}^2\sqrt{q^2}}
  \frac1{(q^2-m_W^2)^2+m_W^2\Gamma_W^2}\frac23\,W_{\rm off-shell}(\theta)
\end{equation}
($g_{w}^2=8m_W^2G_F/\sqrt2=0.4265$) where the polar angle decay distribution
reads
\begin{equation}\label{offshell0}
W_{\rm off-shell}(\theta)
  =\,\frac32\Big(-g^{\mu \mu'}+\frac{q^\mu q^{\mu'}}{m_W^2}\Big)
  \Big(-g^{\nu \nu'}+\frac{q^{\nu}q^{\nu'}}{m_W^2}\Big)
  \rho_{\mu\nu}H_{\mu'\nu'},
\end{equation}
and where $|\vec p_W|=\lambda^{1/2}(m_H^2,m_W^2,q^2)\,/2m_H$ and 
$|\vec p|=\sqrt{q^2}\lambda^{1/2}(1,\mu_1,\mu_2)/2$ are the magnitudes
of the momentum of the $W$ in the $H$ rest system and the momentum of the
quarks in the $W^{\ast+}$ rest system, respectively.

We use the unitary gauge for the electroweak sector in which the numerator of
the gauge boson propagator takes the unitary form written down in
Eq.~(\ref{offshell0}). An identical result is obtained in a general
(`t~Hooft--Feynman) $R_\xi$ gauge where one has to consider also Goldstone
boson exchange. The issue of the gauge invariance of using the Breit--Wigner
form for the propagator numerator has been discussed in
Refs.~\cite{Denner:2006ic,Denner:1999gp}. The gauge invariant complex mass
scheme features such a Breit--Wigner form for the propagator denominator. In
addition, complex masses have to be used in the coupling factors of the $HWW$
and $HZZ$ vertices (see e.g.\ Eq.~(\ref{HWW})) as well as in the relation
between the weak mixing angle $\theta_{W}$ and the gauge boson masses.
Numerically, these corrections to observable quantities amount to less than
one promille and are therefore not discussed any further.

In Eq.~(\ref{hdecay}) we have integrated out a trivial azimuthal angle
dependence. The polarization  of the $W^{\ast+}$ is encoded in the density
matrix function $\rho_{\mu\nu}$ which in turn is determined from the decay
$H\to W^-W^{\ast+}$. The hadron tensor $H_{\mu\nu}$ contains the decay dynamics
of the decay $W^{\ast+}\to q_1\bar q_2$ as described in Sec.~3.

One can separate the spin~1 and spin~0 parts of the propagators in
Eq.~(\ref{offshell0}) by writing\footnote{In the analysis of 
Refs.~\cite{Gao:2010qx,Bolognesi:2012mm} only the spin~1 piece of the
propagator is kept which is adequate for the zero lepton mass case.}  
\begin{equation}\label{spin01}
\Big(-g^{\mu\mu'}+\frac{q^\mu q^{\mu'}}{m_W^2}\Big)
  =\Big(-g^{\mu\mu'}+\frac{q^\mu q^{\mu'}}{q^2}
  -\frac{q^\mu q^{\mu'}}{q^2}(1-\frac{q^2}{m_W^2})\Big).
\end{equation}
Note that, in the product of the two off-shell propagators in 
Eq.~(\ref{hdecay}), the scalar--longitudi\-nal interference term acquires an
extra minus sign.

The polar angle decay distribution of a spin 1 boson decaying into a quark
pair described in Sec.~3 will be augmented by the contribution of a
scalar--longitudinal interference term and a scalar contribution. One has
\begin{eqnarray}\label{offshell}
W_{\rm off-shell}(\theta)
  &=&\frac 32\sum_{m,m'=0,\pm}\rho_{mm}\,d^1_{mm'}(\theta)\,
  d^1_{mm'}(\theta)\,\,H_{m'm'}\nonumber\\&&
  -\frac 32\,\Big(1-\frac{q^2}{m_W^2}\Big)\left(\rho_{t0}H_{t0}
  +\rho_{0t}H_{0t}\right)\cos\theta
  +\frac 32\,\Big(1-\frac{q^2}{m_W^2}\Big)^2\rho_{tt}H_{tt}.\qquad
\end{eqnarray}

In the next step we calculate the density matrix elements of the off-shell
$W^{\ast+}$ in the decay $H \to W^-W^{\ast+}(\uparrow)$ where we sum over the
three polarization states of the on-shell $W^-$. In the SM the Higgs particle
couples to a pair of $W$ bosons via the metric tensor, i.e.\ the matrix
element for $H\to W^-W^+$ is given by 
\begin{equation}\label{HWW}
{\cal M}=im_Wg_{w}\,g_{\mu\nu}\varepsilon^{\ast\mu}_{W^-}
  \varepsilon^{\ast\nu}_q,
\end{equation}
where $\varepsilon_{W^-}$ and $\varepsilon_q$ denote the polarization vectors
of the on-shell $W^-$ and the off-shell $W^{\ast+}$ boson, respectively. On
squaring and summing over the three spin states of the on-shell $W^-$ one 
obtains the density matrix elements 
\begin{equation}\label{denmatr}
\rho_{mm'}=m_W^2\left(-g^{\mu\nu}+\frac{p_W^\mu p_W^\nu}{m_W^2}\right)
  \varepsilon^\ast_{q\mu}(m)\varepsilon_{q\nu}(m').
\end{equation}
The square of the coupling factor $g_{w}$ does not appear in 
Eq.~(\ref{denmatr})
since we have taken the freedom to absorb $g_{w}^2$ in the overall factor in the
rate formula~(\ref{hdecay}).

We calculate the density matrix elements $\rho_{mm'}$ in the Higgs rest frame
with the $z$ axis along the $W^{\ast+}$ momentum $q=p_H-p_W$. Let us collect
the relevant expressions for the four-momentum and the polarization vectors of
the $W^{\ast+}$ boson. One has
\begin{eqnarray}
q^\mu&=&\Big(q_0;0,0,|\vec p_W|\Big),\quad
q_0\ =\ \frac1{2m_H}(m_H^2+q^2-m_W^2),\quad
\varepsilon_q^\mu(\pm)\ =\ \frac1{\sqrt2}\,\Big(0;\mp 1,-i,0\Big),
  \nonumber\\
\varepsilon_q^\mu(0)&=&\frac1{\sqrt{q^2}}\Big(|\vec p_W|;0,0,q_{0}\Big),\qquad
\varepsilon_{q}^\mu(t)\ =\ \frac{q^\mu}{\sqrt{q^2}}
  \ =\ \frac1{\sqrt{q^2}}\Big(q_{0};0,0,|\vec{p}_W|\Big).
\end{eqnarray}

The propagation of the scalar degree of freedom can be made explicit by
expanding the propagator in terms of a complete set of polarization vectors
(see e.g.\ Ref.~\cite{Kadeer:2005aq,Korner:1989qb})
\begin{equation}
-g_{\mu\nu}+\frac{q_{\mu}q_{\nu}}{m^{2}_{W}}
=-\sum_{m,m'=t,\pm,0}\eps_{q\mu}(m)\eps^{\ast}_{q\nu}(m')g_{mm'}
\end{equation}
where $g_{mm'}=\mbox{diag\,}\{A;-1,-1,-1\}$ with $A=(1-q^{2}/m^{2}_{W})$. The
scalar degree of freedom proportional to $\eps_{q\mu}(t)\eps^{\ast}_{q\nu}(t)$
propagates from the $HWW$ vertex to the $Wf\bar f$ vertex. The scalar degree
of freedom only comes into play for nonzero fermion masses.

On evaluating Eq.~(\ref{denmatr}) one obtains
\begin{eqnarray}\label{dmwstar}
\rho_{++}\ =\ \rho_{--}&=&m_W^2,\qquad\qquad\qquad\qquad 
\rho_{00}\ =\ m_W^2\Big(1+\frac{m_H^2}{q^2m_W^2}|\vec p_W|^2\Big),\nonumber\\
\rho_{0t}\ =\ \rho_{t0}&=&m_W^2\frac{m_H|\vec p_W|}{2m_W^2q^2}
  \left(m_H^2-m_W^2-q^2\right),\qquad
\rho_{tt}\ =\ m_W^2\frac{m_H^2}{q^2m_W^2}|\vec{p}_W|^2.\qquad
\end{eqnarray}
At threshold (maximal recoil) when $q^2\to(m_1+m_2)^2$, and for 
$m_i\to 0$, the longitudinal and scalar contributions 
$\rho_{00}=\rho_{tt}=\rho_{t0}=(m_H^2-m_W^2)^2/4q^2$ become dominant. On the
other end of the $q^2$ spectrum (zero recoil) where $|\vec p_W|=0$, one finds 
$\rho_{++}=\rho_{00}=\rho_{--}=m_W^2$ and $\rho_{tt}=\rho_{t0}=0$. 

Since the decay $H\to W^-W^{\ast+}$ is parity-conserving, the transverse 
density matrix elements $\rho_{++}$ and $\rho_{--}$ are identical to each 
other, i.e.\ one has $\rho_{++}-\rho_{--}=0$. This means that there is no
parity-violating contribution to the $\cos\theta$ coefficient in the (first)
spin~1 part of Eq.~(\ref{offshell}) (see Eq.~(\ref{Wtheta})). The second
$\cos\theta$ contribution in Eq.~(\ref{offshell}) does not have a
parity-violating origin but is a parity-odd effect. It arises from the 
scalar--longitudinal interference contribution with $J^P$ properties 
$(0^+,1^-)$ (VV) and $(0^-,1^+)$ (AA), resulting in a parity-odd contribution.

The polarized decay functions $H_{\pm\pm}$ and $H_{00}$ have been calculated
before. The LO and NLO forms of the additional polarized decay functions
$H_{tt}$ and $H_{t0}$ can be found in Sec.~2 and in Appendix~D. For the
convenience of the reader we list $H_{tt}$ and $H_{t0}$ together with their
$O(\mu_i)$ mass expansion. One has
\begin{eqnarray}
\label{Htt}
H_{tt}&=&4N_cq^2\left(1-\mu_1-\mu_2-\lambda+H^1_S(\alpha_s)\right)\nonumber\\
  &=&4N_cq^2\left(\mu_1+\mu_2+\ldots+\frac{\alpha_s}{6\pi} 
  \Big(18\mu_1+18\mu_2+12\mu_1\ln\mu_1+12\mu_2\ln\mu_2+\ldots\Big)\right)\qquad
\end{eqnarray}
and
\begin{eqnarray}\label{Ht0}
H_{t0}&=&H_{0t}\ =\ 4N_cq^2\left(-(\mu_1-\mu_2)\sla+H^1_{0t}(\alpha_s)\right)
  \nonumber\\
  &=&-4N_cq^2\bigg(\mu_1-\mu_2+\ldots\nonumber\\&&\strut
  +\frac{\alpha_{s}}{6\pi}\Big(26\mu_1-14\mu_2-4\pi(\mu_1-\mu_2)
  +12\mu_1\ln\mu_1+12\mu_2\ln\mu_2+\ldots\Big)\bigg).
\end{eqnarray} 

On integrating Eq.~(\ref{hdecay}) over $\cos\theta$ one obtains the 
differential $q^2$ rate which is given by 
\begin{eqnarray}\label{hdecay1}
\frac{d\Gamma}{dq^2}&=&\frac{g_{w}^4}{1024\pi^3}|V_{12}|^2
\frac{|\vec p_W||\vec p\,|}{m_H^2\sqrt{q^2}}
  \frac1{(q^2-m_W^2)^2+m_W^2\Gamma_W^2}\nonumber\\&&\strut\times
  \frac23\Big((\rho_{++}+\rho_{00}+\rho_{--})(H_{++}+H_{00}+H_{--})
  +3(1-\frac{q^2}{m_W^2})^2\rho_{tt}H_{tt}\Big).
\end{eqnarray} 
In the zero quark mass limit $m_{i}\to 0$ where
$H_{++}+H_{00}+H_{--}=8N_cq^2$ and $H_{tt}=0$, the Born-term rate calculated
from Eq.~(\ref{hdecay1}) can be seen to agree with the result of
Refs.~\cite{Keung:1984hn,Djouadi:2005gi} when $N_c$ is set to 
one.\footnote{As pointed out in Ref.~\cite{Keung:1984hn}, the corresponding
result in Ref.~\cite{Rizzo:1980gz} is too small by a factor of $3/4$.}

\begin{figure}
\begin{center}
\epsfig{figure=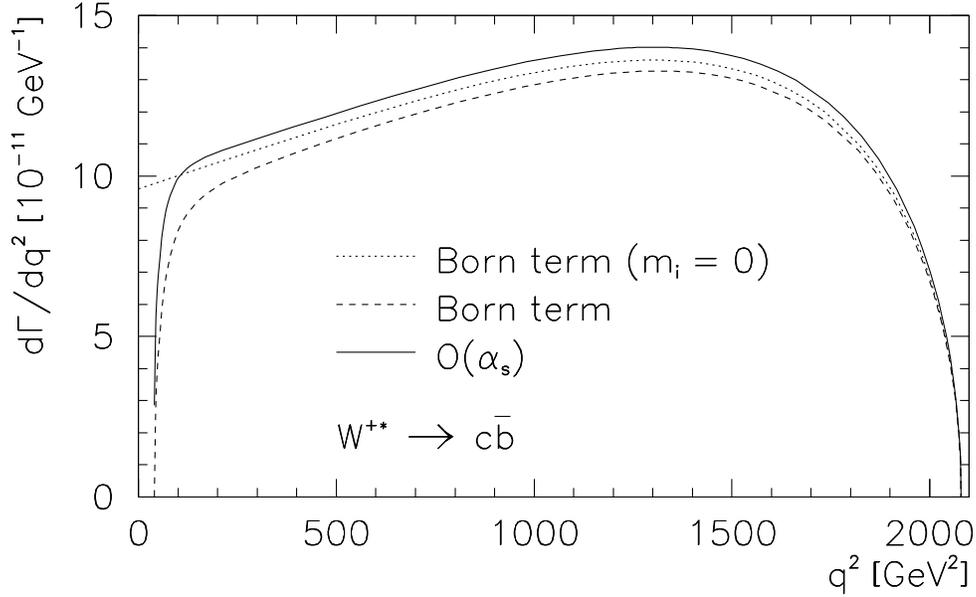, scale=0.8}
\end{center}
\caption{\label{dgamh}Differential rate for the three body decay
  $H\to W^-+W^{\ast+}(\to c\bar b)$. The three curves correspond to
  (i) Born term $(m_i=0)$ (dotted line)
  (ii) Born term $(m_i\neq 0)$ (dashed line) and
  (iii) $O(\alpha_s)$ with $(m_i\neq 0)$ (full line).}
\end{figure}

In our numerical discussion we again concentrate on the mode
$H\to W^-+W^{\ast+}(\to c\bar b)$ in order to highlight quark mass effects
even if this mode is suppressed by $|V_{cb}|^2=(0.041)^2$. In Fig.~\ref{dgamh}
we show the $q^2$ dependence of the rate. Let us begin our discussion with the
Born-term contributions. In the threshold region, where the longitudinal
$W^{\ast+}$ dominates, the $m_i\neq 0$ differential rate clearly shows the
appropriate threshold behaviour
$2|\vec p\,|/\sqrt{q^2}=\lambda^{1/2}(1,\mu_1,\mu_2)$, i.e.\ the differential
rate vanishes at threshold. This vanishing is not seen for the
$m_i=0$ curve. This can be understood by taking the $m_i\to 0$ limit of
$\lambda^{1/2}(1,\mu_1,\mu_2)$ keeping $q^2$ small and fixed with
the result $\lambda^{1/2}(1,\mu_1,\mu_2)\,\to 1$. For the $q^2=0$ value of
the differential $m_i=0$ rate one then obtains
\begin{equation}\label{Wq^2zero}
\frac{d\Gamma}{dq^2}\,\bigg|_{q^2=0}=\frac{g_{w}^4}{1024\pi^3}|V_{12}|^2
  \frac{N_{c}}3\frac{(m_H^2-m_W^2)^3}{m_H^3m_W^2(m_W^2+\Gamma_W^2)}
  =9.554\cdot 10^{-11}\GeV^{-1}.
\end{equation}
in agreement with Fig.~\ref{dgamh}. At higher values of $q^2$ the difference
between the $m_i=0$ and $m_i\neq 0$ Born-term curves becomes smaller and
smaller. The radiative corrections are largest in the threshold region. Away
from the threshold region they amount to over $10\,\%$ and are thus
considerably larger than what would result from the simple estimate
$\alpha_s/\pi\sim 3.7\,\%$. We mention that the radiative corrections to the
LO $m_i=0$ curve in Fig.~\ref{dgamh} is simply given by multiplying the LO
result by $(1+\alpha_s/\pi)$.

Fig.~\ref{dgamh} also shows that the $O(\alpha_s)$ $m_i\neq 0$ rate does not
go to zero at threshold. This can be traced to the presence of the NLO
chromodynamic Coulomb singularity at threshold. The Coulomb singularity
proportional to $\lambda^{-1/2}$ (see Eq.~(\ref{coulomb})) is cancelled by the
overall rate factor $|\vec p\,|=\sqrt{q^2}\lambda^{1/2}/2$ resulting in a
finite contribution at threshold proportional to $\alpha_s$. One can estimate
the finite threshold value of the $O(\alpha_s)$ rate by neglecting terms of
$O(q^2/m_W^2)$ in Eq.~(\ref{hdecay1}) whence one can express the finite
threshold value in terms of the LO $m_i=0$ contribution in
Eq.~(\ref{Wq^2zero}). One then obtains
\begin{equation}\label{coulombW}
\frac{d\Gamma}{dq^2}\,\bigg|_{\rm thresh}\approx\alpha_s\,
\frac{32\pi}3\,\mu_1\mu_2\,\frac{d\Gamma}{dq^2}\,\bigg|_{q^2=0}.
\end{equation}
Using $\alpha_s(q^2=(4.8+1.5)^2\GeV^2)=0.165$, one obtains approximate
agreement with Fig.~\ref{dgamh}. As has been emphasized before, perturbation
theory cannot be trusted in the threshold region and therefore the treatment
of the decay $W^{\ast+}\to c\bar b$ requires a nonperturbative treatment
including a resummation of the chromodynamic Coulomb singularity. The above
exercise leading to Eq.~(\ref{coulombW}) merely serves to check on the
consistency of our calculation.

\begin{figure}
\begin{center}
\epsfig{figure=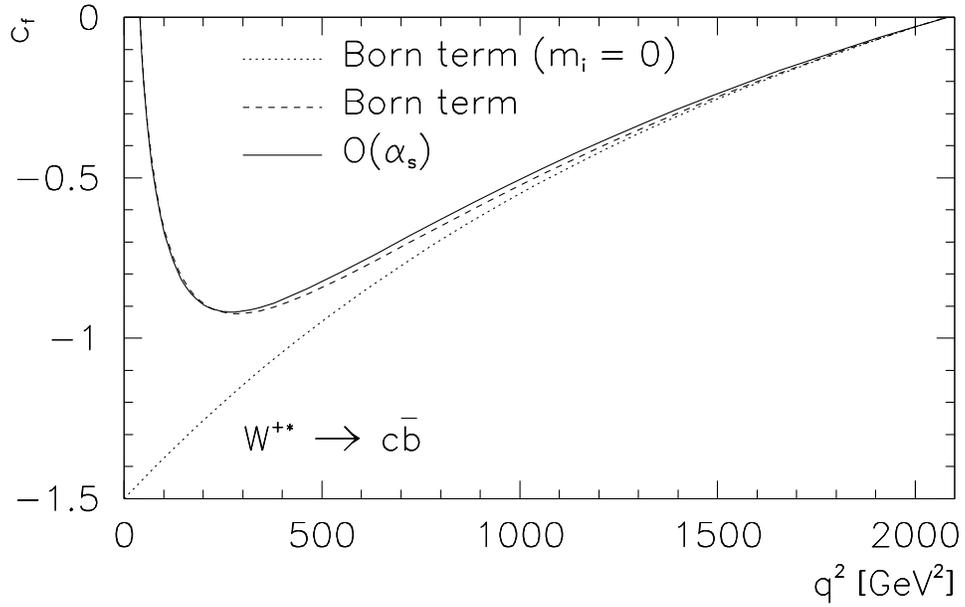, scale=0.8}
\end{center}
\caption{\label{dgamhc}Convexity parameter $c_f(q^2)$ as a function of $q^2$.
  Labelling of curves as in Fig.~\ref{dgamh}}
\end{figure}

In Fig.~\ref{dgamhc} we show a plot of the $q^2$ dependence of the convexity
parameter. The convexity parameter is obtained from Eq.~(\ref{hdecay1}) by
replacing $(\rho_{++}+\rho_{00}+\rho_{--})(H_{++}+H_{00}+H_{--})$ by
$3/4(\rho_{++}-2\rho_{00}+\rho_{--})(H_{++}-2H_{00}+H_{--})$, setting the
scalar contribution to zero, and then dividing by the differential
rate~(\ref{hdecay1}). At threshold and at zero recoil the convexity parameter
can be seen to go to zero at both ends of the $q^2$ spectrum because one has
$H_{++}-2H_{00}+H_{--}\to 0$ at threshold and
$\rho_{++}-2\rho_{00}+\rho_{--}\to 0$ at zero recoil.

An interesting exercise is to calculate the LO convexity parameter in the
threshold region. Neglecting terms of $O(q^2/m_Z^2)$, as before, one obtains
\begin{equation}\label{cfWW}
c_f\sim-\frac 32\,\pfrac\lambda{3-3\mu_1-3\mu_2-2\lambda}.
\end{equation}
The expression~(\ref{cfWW}) has the correct threshold behaviour. Keeping
$q^2$ fixed (and small), and taking the limit $m_i\to 0$ one has $\mu_i\to 0$,
$\lambda\to 1$ and one obtains $c_f=-3/2$ in agreement with Fig.~\ref{dgamhc}.
\begin{figure}
\begin{center}
\epsfig{figure=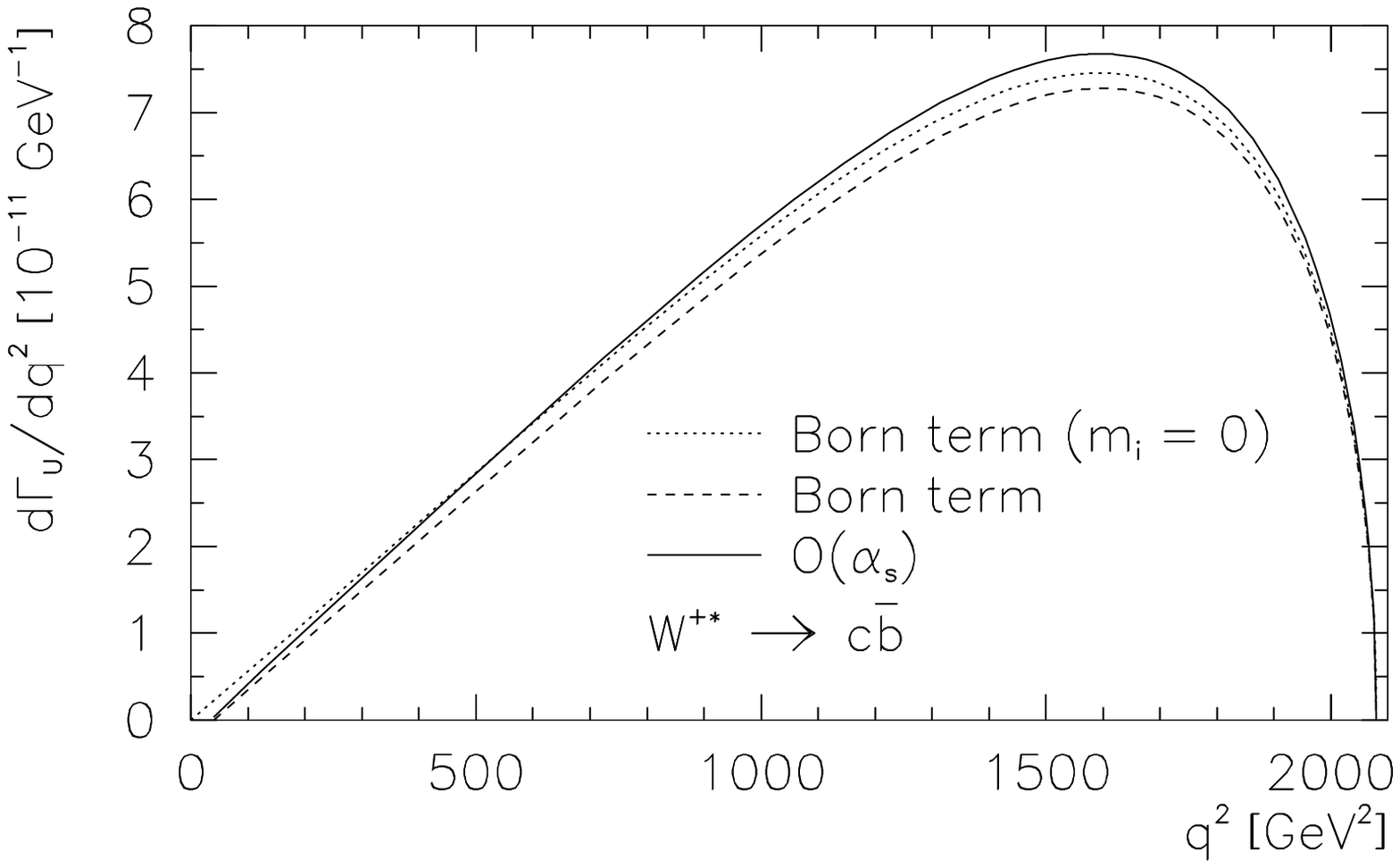, scale=0.8}
\end{center}
\caption{\label{dgamhu} Differential rate $d\Gamma_{U}/d{q^2}$.
  Labelling of curves as in Fig.~\ref{dgamh}}
\end{figure}  
\begin{figure}
\begin{center}
\epsfig{figure=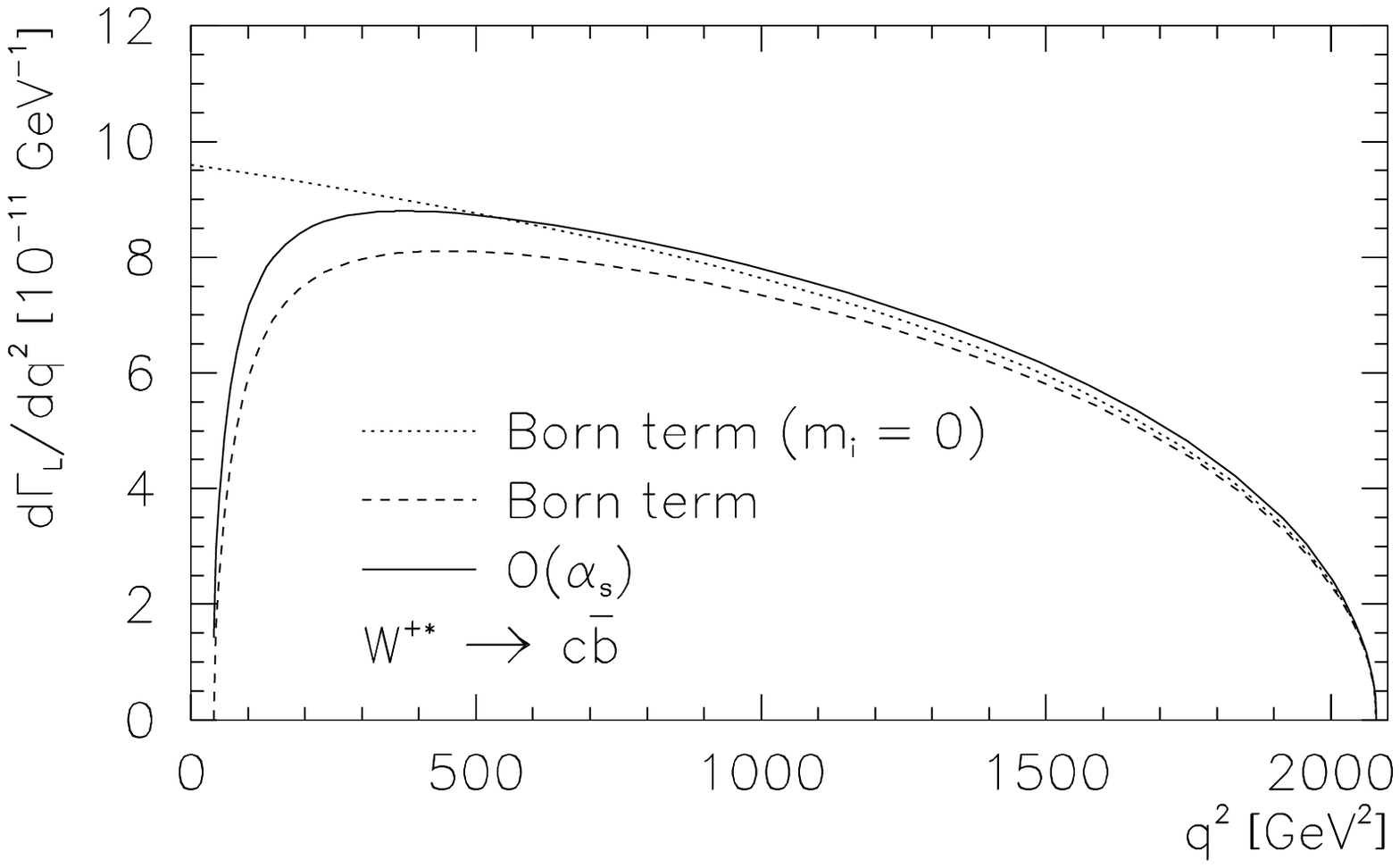, scale=0.8}
\end{center}
\caption{\label{dgamhl} Differential rate $d\Gamma_{L}/d{q^2}$. 
  Labelling of curves as in Fig.~\ref{dgamh}}
\end{figure}  
\begin{figure}
\begin{center}
\epsfig{figure=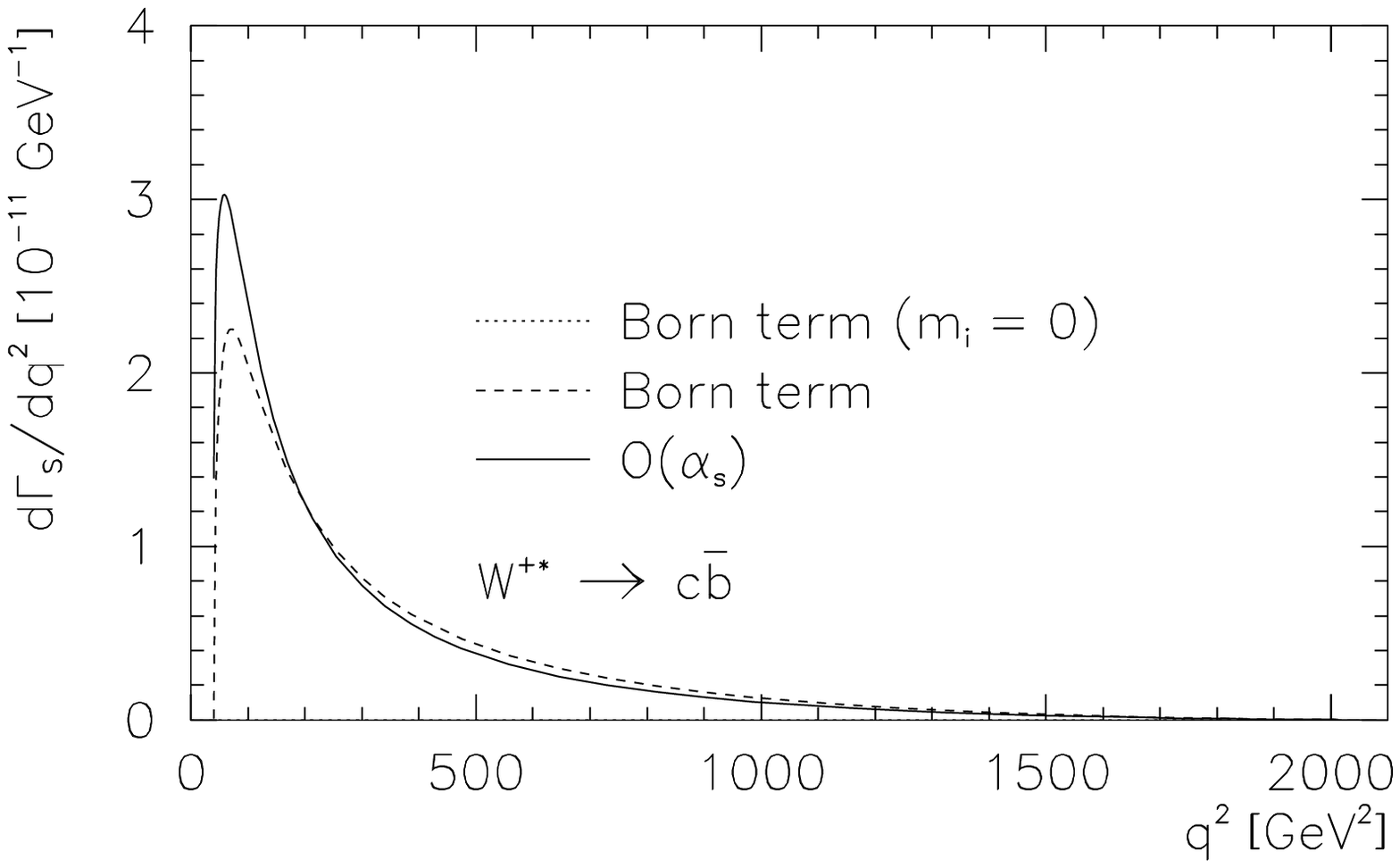, scale=0.8}
\end{center}
\caption{\label{dgamhs} Differential rate $d\Gamma_{S}/d{q^2}$. 
  Labelling of curves as in Fig.~\ref{dgamh}}
\end{figure}

In Figs.~\ref{dgamhu}--\ref{dgamhs} we decompose the total differential rate
$d\Gamma/dq^2$ in terms of the three partial unpolarized transverse ($U$), 
longitudinal ($L$) and scalar ($S$) contributions $d\Gamma_U/dq^2$,
$d\Gamma_L/dq^2$ and $d\Gamma_S/dq^2$, where the three partial rates are
defined by the contributions of the density matrix elements
$\rho_{++}+\rho_{--}$, $\rho_{00}$ and $\rho_{tt}$, respectively. The total
rate is then given by
$d\Gamma/dq^2=d\Gamma_U/dq^2+d\Gamma_L/dq^2+d\Gamma_S/dq^2$.

Fig.~\ref{dgamhu} shows that the transverse rate is weighted toward higher
$q^2$ values, whereas the longitudinal rate is more evenly distributed
(Fig.~\ref{dgamhl}). The scalar rate is considerably smaller and shows a peak
close to threshold (Fig.~\ref{dgamhs}). The peak value is strongly enhanced by
the radiative corrections. The radiative corrections to the transverse rate
are small. The radiative corrections to the longitudinal rate can be seen to
be quite pronounced close to threshold which, in part, is due to the increase
of $\alpha_s$ due to running.

\begin{table}[ht]\begin{center}
\begin{tabular}{|l||c|c|c|c|}\hline
&Born\ $m_{i}=0$ &Born\ $m_{i}\neq 0$
&$O(\alpha_{s})\ m_{i}=0$&$O(\alpha_{s})\ m_{i}\neq 0$\\\hline
$W^{\ast+}\to c\bar b$&&&&\\
$\Gamma$\phantom{xxxx}&$2.43\cdot 10^{-7}$&$2.30\cdot 10^{-7}$
&$2.55\cdot 10^{-7}$&$2.45\cdot 10^{-7}$\\[-2.0ex]
$\Gamma_{U}$&$9.79\cdot 10^{-8}$&$9.40\cdot 10^{-8}$
&$1.02\cdot 10^{-7}$&$9.98\cdot 10^{-8}$\\[-2.0ex]
$\Gamma_{L}$&$1.45\cdot 10^{-7}$&$1.29\cdot 10^{-7}$
&$1.52\cdot 10^{-7}$&$1.39\cdot 10^{-7}$\\[-2.0ex]
$\Gamma_{S}$&$0$&$6.67\cdot 10^{-9}$&$0$&$6.81\cdot 10^{-9}$\\[-2.0ex]
$A_{FB}$&$0$&$0.0194$&$0$&$0.0190$\\
\hline
$W^{\ast+}\to c\bar s$&&&&\\
$\Gamma$\phantom{xxxx}&$1.37\cdot 10^{-4}$&$1.36\cdot 10^{-4}$
&$1.44\cdot 10^{-4}$&$1.43\cdot 10^{-4}$
\\[-2.0ex]
$\Gamma_{U}$&$5.52\cdot 10^{-5}$&$5.50\cdot 10^{-5}$
&$5.77\cdot 10^{-5}$&$5.76\cdot 10^{-5}$\\[-2.0ex]
$\Gamma_{L}$&$8.19\cdot 10^{-5}$&$8.06\cdot 10^{-5}$
&$8.58\cdot 10^{-5}$&$8.49\cdot 10^{-5}$\\[-2.0ex]
$\Gamma_{S}$&$0$&$7.46\cdot 10^{-7}$&$0$&$6.53\cdot 10^{-7}$\\[-2.0ex]
$A_{FB}$&$0$&$-0.00433$&$0$&$-0.00339$\\
\hline
$Z^{\ast}\to b \bar b$&&&&\\
$\Gamma$&$7.47\cdot 10^{-6}$&$5.98\cdot 10^{-6}$
&$7.82\cdot 10^{-6}$&$6.68\cdot 10^{-6}$\\[-2.0ex]
$\Gamma_{U}$&$3.03\cdot 10^{-6}$&$2.51\cdot 10^{-6}$
&$3.16\cdot 10^{-6}$&$2.77\cdot 10^{-6}$\\[-2.0ex]
$\Gamma_{L}$&$4.44\cdot 10^{-6}$&$2.95\cdot 10^{-6}$
&$4.66\cdot 10^{-6}$&$3.34\cdot 10^{-6}$\\[-2.0ex]
$\Gamma_{S}$&$0$&$5.11\cdot 10^{-7}$&$0$&$5.72\cdot 10^{-7}$\\[-2.0ex]
$A_{FB}$&$0$&$0$&$0$&$0.000554$\\
\hline
$Z^{\ast}\to c\bar c$&&&&\\
$\Gamma$&$5.79\cdot 10^{-6}$&$5.65\cdot 10^{-6}$
&$6.06\cdot 10^{-6}$&$5.99\cdot 10^{-6}$\\[-2.0ex]
$\Gamma_{U}$&$2.35\cdot 10^{-6}$&$2.29\cdot 10^{-6}$
&$2.45\cdot 10^{-6}$&$2.42\cdot 10^{-6}$\\[-2.0ex]
$\Gamma_{L}$&$3.45\cdot 10^{-6}$&$3.20\cdot 10^{-6}$
&$3.61\cdot 10^{-6}$&$3.42\cdot 10^{-6}$\\[-2.0ex]
$\Gamma_{S}$&$0$&$1.55\cdot 10^{-7}$&$0$&$1.42\cdot 10^{-7}$\\[-2.0ex]
$A_{FB}$&$0$&$0$&$0$&$0.000424$ \\
\hline
\end{tabular}
\caption{\label{tab2}Integrated rates $\Gamma,\Gamma_U,\Gamma_L,\Gamma_S$ and
forward--backward asymmetry $A_{FB}$ for $H\to W^-+W^{\ast+}(\to c\bar b)$,
$H\to W^-+W^{\ast+}(\to c\bar c)$, $H\to Z+Z^\ast(\to b\bar b)$ and
$H\to Z+Z^\ast(\to c\bar c)$. All entries are given in units of $\GeV$ except
for $A_{FB}$.}
\end{center}\end{table}

In Tab.~\ref{tab2} we present our numerical results for the integrated total
rate and the integrated partial rates for $W^+\to c\bar b$. One can see that
the integrated longitudinal rate $\Gamma_L$ slightly dominates over the
integrated transverse rate $\Gamma_U$. The scalar rate $\Gamma_S$ is quite
small and contributes to the total rate at the $2.9\,\%$ level. The LO total
rate is reduced by $5.8\,\%$ through mass effects where the biggest reduction
comes from the longitudinal rate ($11.7\,\%$). Radiative corrections increase
the LO rates by $6.2\,\%-7.8\,\%$ except for the scalar rate which is
increased only by $2.1\,\%$. We also list the value of the forward--backward
asymmetry $A_{FB}$ which, as has been discussed before, is a parity-odd effect
contributed to by the parity-conserving scalar--longitudinal interference term.
The forward--backward asymmetry is positive (see Eqs.~(\ref{offshell})
and~(\ref{Ht0})) and receives its main contribution from the region close to
threshold. $A_{FB}$ is of the same order of magnitude as $\Gamma_S/\Gamma_L$.
For comparison, in Tab.~\ref{tab2} we also include results for the process
$W^+\to c\bar s$ ($|V_{cs}|=0.97345\pm 00016$~\cite{Nakamura:2010zzi}).

\begin{figure}
\begin{center}
\epsfig{figure=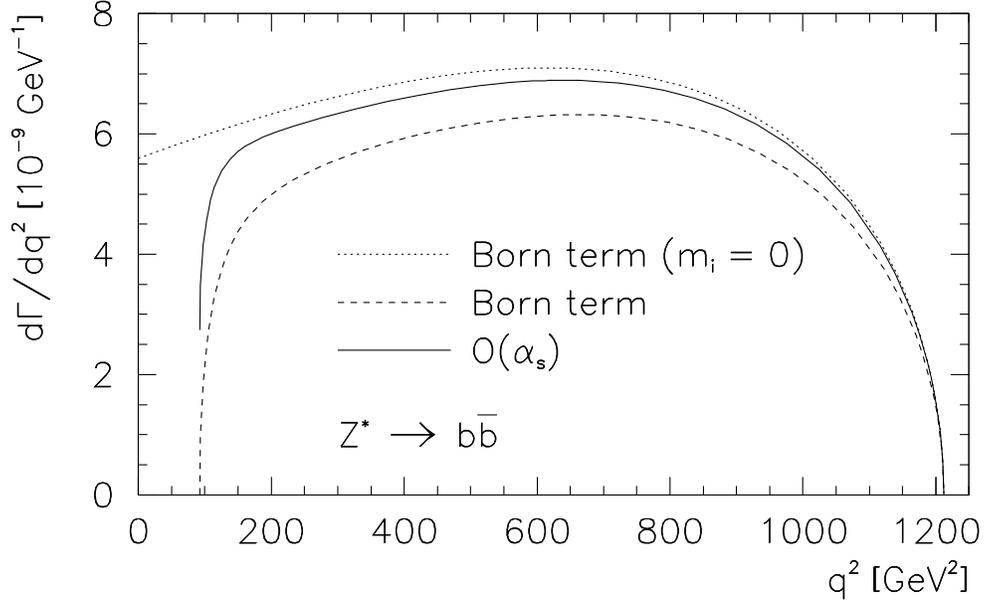, scale=0.8}
\end{center}
\caption{\label{dgamhzb} Differential rate for the three body decay
  $H\to Z+Z^{\ast}(\to b+\bar b)$. Labelling of curves as in Fig.~\ref{dgamh}.}
\end{figure}

Quark mass effects can be expected to play a larger role in e.g.\ the decay
$H\to Z+Z^\ast(\to b\bar b)$. First, the $b\bar b$ threshold is higher than
the $c\bar b$ threshold, and second, the phase space is reduced due to the
larger mass of the $Z$ boson, i.e.\ the physical $q^2$ range becomes smaller.
An extra bonus is the fact that the decay $Z^\ast\to b\bar b$ is not CKM
suppressed. For the differential decay distribution one obtains
($\sin^2\theta_W=0.23188$)
\begin{eqnarray}\label{hdecay2}
\frac{d\Gamma}{dq^2}&=&\frac 12\,\frac{g_{w}^4}{1024\pi^3}\frac1{\cos^4\theta_W}
  \frac{|\vec p_Z||\vec p\,|}{m_H^2\sqrt{q^2}}
  \frac1{(q^2-m_Z^2)^2+m_Z^2\Gamma_Z^2}\nonumber\\&&\strut\times
  \frac 23\Big((\rho_{++}+\rho_{00}+\rho_{--})
 \frac12(v_f^2H^{VV}_{U+L}+a_f^2H^{AA}_{U+L})\nonumber\\&&\strut\qquad\qquad
  +3(1-\frac{q^2}{m_Z^2})^2\,\rho_{tt}
  (v_f^2H^{VV}_{tt}+a_f^2H^{AA}_{tt})\Big),
\end{eqnarray}
where the gauge boson momentum now is
$|\vec p_Z|=\sqrt{\lambda(m_H^2,m_Z^2,q^2)}\,/2m_H$, and $\vec p_W,m_W$ are
replaced by $\vec p_Z,m_Z$ in the expressions for $\rho_{mm'}$ in
Eq.~(\ref{dmwstar}). The electroweak coupling coefficients are given by
\begin{eqnarray}
v_f&=&1-\frac 83\sin^2\theta_W,\qquad a_f=1\qquad\mbox{for}\quad u,c,t 
\nonumber \\
v_f&=&-1+\frac 43\sin^2\theta_W,\qquad a_f=-1\qquad\mbox{for}\quad d,s,b.
\end{eqnarray}
In Fig.~\ref{dgamhzb} we provide a plot of the differential $q^2$ rate for
$H\to Z+Z^{\ast}(\to b\bar b)$ where we use $m_Z=91.1876\pm0.0021\GeV$,
$\Gamma_Z=2.4952\pm 0.0023\GeV$~\cite{Nakamura:2010zzi}. Again the
differential LO rate shows the appropriate threshold behaviour for
$m_b\neq 0$, i.e.\ the differential rate vanishes at threshold $q^2=4m_b^2$.

The corresponding $m_b=0$ LO rate shows no apparent vanishing at threshold
for the same reason as in the corresponding $H\to W^-W^{\ast+}$ case. The
differential rate at $q^2=0$ and for $m_b=0$ is given by
\begin{equation}\label{Zq^2zero}
\frac{d\Gamma}{dq^2}\,\bigg|_{q^2=0}=\frac12\,\frac{g_{w}^4}{1024\pi^3}
  \frac1{\cos^4\theta_W}\frac{N_c}3
  \frac{(m_H^2-m_Z^2)^3}{m_H^3m_Z^2(m_Z^2+\Gamma_Z^2)}
  \frac12\frac{v_f^2+a_f^2}2=0.56\cdot 10^{-8}\GeV^{-1}.
\end{equation}
in agreement with Fig.~\ref{dgamhzb}. 

As Fig.~\ref{dgamhzb} shows, the NLO rate does not go to zero at threshold.
As in the charged current case this can be traced to the presence of the NLO
chromodynamic Coulomb singularity at threshold. One can estimate the finite
threshold value of the $O(\alpha_s)$ rate by neglecting terms of $O(q^2/m_Z^2)$
in Eq.~(\ref{hdecay2}) whence one can express the finite threshold value in
terms of the LO $m_b=0$ contribution in Eq.~(\ref{Zq^2zero}). One then obtains
\begin{equation}
\frac{d\Gamma}{dq^2}\,\bigg|_{\rm thresh}\approx\,\alpha_s\,
  \frac{(v_f^2+a_f^2)}{(v_f^2+3a_f^2)}\frac{16\pi}{3}\mu^2\,
  \frac{d\Gamma}{dq^2}\,\bigg|_{q^2=0}.
\end{equation}
Note that the contribution proportional to $3a_f^2$ results from the scalar
contribution in Eq.~(\ref{hdecay2}). By a visual inspection of
Fig.~\ref{dgamhzb}, the approximation can be seen to be quite good. 
Similar to the calculation leading up to Eq.~(\ref{cfWW}) one can calculate
the LO convexity parameter in the threshold region. Neglecting again terms of
$O(q^2/m_{Z}^2)$ one finds
\begin{equation}\label{cfZZ}
c_f=-\frac 32\pfrac{1-4\mu}{1+2\mu}
\end{equation}
which is just the limiting case of Eq.~(\ref{cfWW}) for $\mu_1=\mu_2:=\mu$.
Curiously the intricate dependence on the electroweak coupling parameters
$c_f$ and $a_f$ has dropped out when taking the ratio. In the mass-zero case
and at $q^2=0$ one has exactly $c_f=-3/2$.

\begin{figure}
\begin{center}
\epsfig{figure=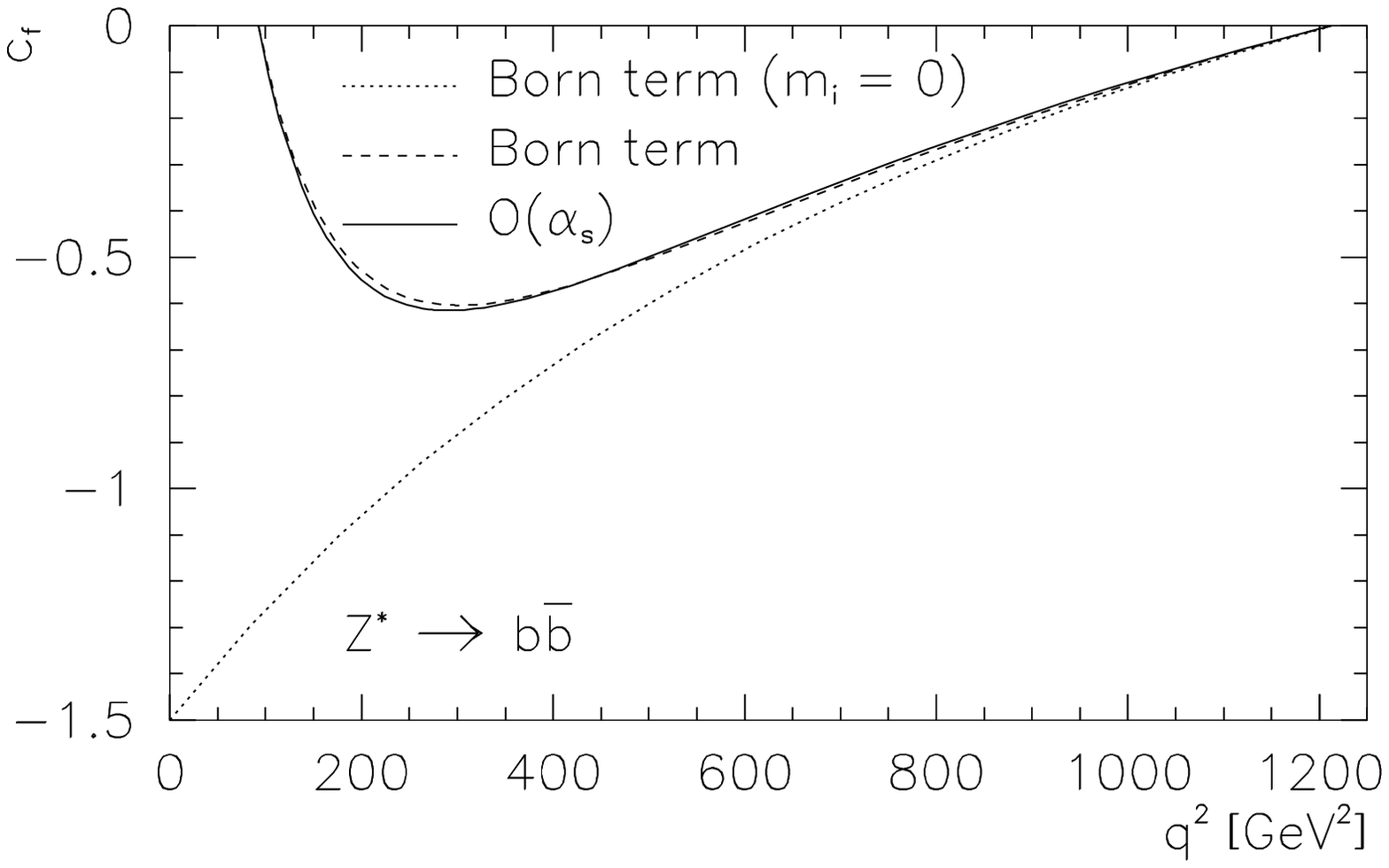, scale=0.8}
\end{center}
\caption{\label{dgamhczb}$q^2$ dependence of the convexity parameter $c_f(q^2)$
   for $H\to Z+Z^{\ast}(\to b+\bar b)$. Labelling of curves as in
Fig.~\ref{dgamh}}
\end{figure}
  
\begin{figure}
\begin{center}
\epsfig{figure=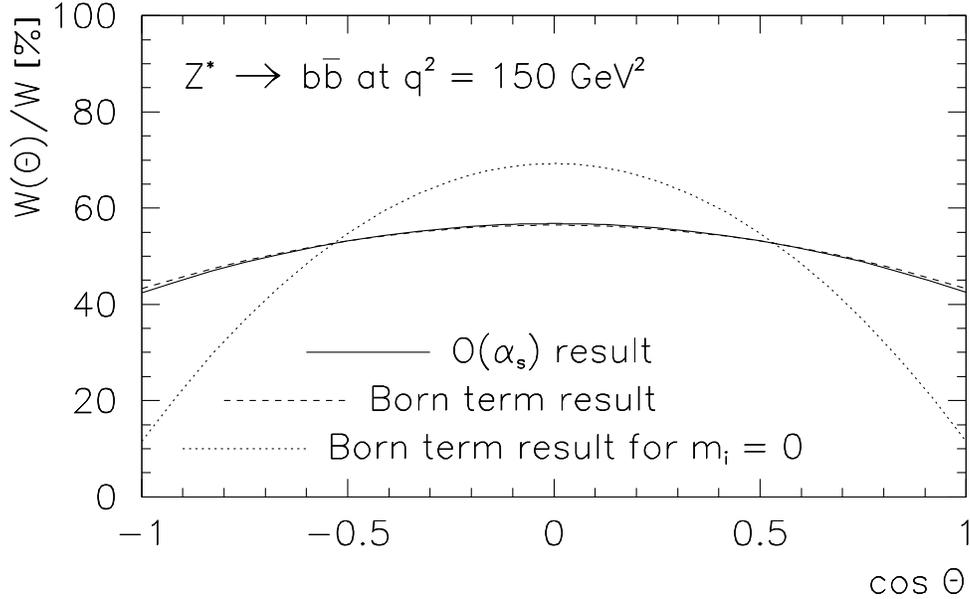, scale=0.8}
\end{center}
\caption{\label{nurkzbr}Polar angle distribution for $Z^{\ast}(\to b+\bar b)$
  at $q^2=150\GeV^2$. Labelling of curves as in Fig.~\ref{dgamh}}
\end{figure}

In Fig.~\ref{dgamhczb} we show a plot of the $q^2$ dependence of the convexity
parameter $c_f$. In the threshold region the convexity parameter behaves very
differently for $m_b=0$ and $m_b\neq 0$ (the radiative corrections are quite
small). This implies that the polar angle distributions are very different for
the two cases. In order to illustrate this effect we choose $q^2=150\GeV^2$
and, in Fig.~\ref{nurkzbr}, plot the corresponding $\cos\theta$ distribution.
At this value of $q^2$ one is well outside of the nonperturbative threshold
region. Since the convexity parameter is negative (see Eq.~(\ref{cfZZ})), one
has a downward open parabola. Mass effects can be seen to be crucial for the
correct description of the $m\neq 0$ angular decay distribution which is much
flatter than  the $m=0$ distribution. The three curves correspond to convexity
parameters of $c_f=-1.154$ (LO;\,$m_b=0$), $c_f=-0.388$ (LO) and
$c_f=-0.407$ (NLO).

In Table~\ref{tab2} we have listed numerical values for the various integrated
partial rates and for the asymmetry parameter $A_{FB}$ for both
$Z^\ast\to b\bar b$ and for $Z^\ast\to c\bar c$. Quark mass effects and scalar
contributions can be seen to be quite large in particular for the $b\bar b$ 
case. In the $b\bar b$ case, mass effects decrease the LO rate by $20.1\,\%$
where most of this reduction comes from $\Gamma_L$. The scalar contribution
amounts to $8.6\,\%$ of the total contribution. The radiative corrections
increase all four rates by $O(10\,\%)$. The scalar--longitudinal interference
contribution sets in only at $O(\alpha_s)$ since the Born-term contribution to
$H_{0t}$ vanishes, i.e.\ the forward--backward asymmetry is proportional to
$\alpha_s$ and therefore small. This is borne out by the tiny numerical value
of $A_{FB}$ listed in Table~\ref{tab2}. The numbers in Table~\ref{tab2} for
the $c\bar c$ case follow a similar pattern, though quark mass and off-shell
effects are smaller.

In this section we have assumed one gauge boson to be 
on-shell and the other opposite-side gauge boson to be off-shell. The
on-shell approximation can be dropped by also taking the on-shell gauge boson
off its mass shell using, again, a Breit--Wigner form for the propagator. We 
find upward corrections to the rate of $5.2\,\%$ for $H\to WW$ and $19.9\,\%$
for $H\to ZZ$.

In the present calculation we have used a factorized form for
the opposite-side fermion pair decays which is only justified when one does
not have identical fermions in the final state. If one has identical fermions
in the final state as in $H \to Z^{\ast}Z^{\ast}(\to ff \bar f \bar f)$, there 
will be interference effects involving the pairs of identical fermions. In 
order to account for such interference effects, a full-fledged calculation of
$H\to Z^*Z^*$ with subsequent four-body decays is required, as has been done
in Refs.~\cite{Bredenstein:2006rh,Bredenstein:2006ha}. As shown in
Ref.~\cite{Denner:2011mq}, these interference effects can lead to a
substantial reduction in rate. For example, for a $126\GeV$ Higgs boson
interference effects reduce the branching ratio of $H\to ee\mu\mu$ by
$45\,\%$ when going to the decay $H\to eeee$.

\section{Summary and Conclusions}
We have calculated the NLO QCD corrections to the polarized decay functions
in the decay of an off-shell and on-shell polarized $W^+$ gauge boson into 
massive quark--antiquark pairs $W^+(\uparrow)\to q_1\,\bar q_2$, keeping the
quark masses finite. Using these NLO results for the decay process as well as
previous results on the NLO corrections to the production process $t\to b+W^+$
we have studied the NLO corrections to the polar angle decay distribution in
the cascade decay $t\to b+W^+$ followed by $W^+\rightarrow q_1\,\bar q_2$. We
have found that the NLO final-state corrections to the decay distribution are
somewhat larger than the NLO initial state corrections. Altogether we find
that the NLO corrections lead to a flatter angular decay distribution
$W(\theta)$.

The decay analysis was done in the $W^+$ rest frame which has the maximal
sensitivity to $W^+$ polarization effects. Polarization effects of the $W^+$
boson will be visible also in other reference frames such as the laboratory
frame. It is therefore always important to retain $W^+$ polarization effects
in radiative correction calculations (see e.g.\
Refs.~\cite{Bredenstein:2006rh,Bredenstein:2006ha,Jager:2013mu,Jager:2012xk,%
Oleari:2003tc,Melia:2011dw}).

We have presented our results in a general form involving the spin~0 and
spin~1 pieces of the $(VV)$, $(AA)$, $(VA)$ and $(AV)$ current contributions 
separately. Our results can thus also be applied to on-shell $Z$ decays
and off-shell $Z^\ast$ decays (as in Sec.~9) and also to extensions of the SM.

In this paper we have discussed the decays $W^+\to q_1\,\bar q_2$ of
positively charged $W^+$ bosons. The corresponding results for negatively
charged bosons $W^-\to\bar q_1\,q_2$ can be obtained from the CP invariance of
the interaction. One finds~\cite{Groote:2012xr}
\begin{equation}\label{wminus}
H_{mm}(W^-\to\bar{q}_1q_2;\mu_1,\mu_2;z'\parallel q_2)
  =H_{mm}(W^+\to q_1\bar{q}_2;\mu_2,\mu_1;z'\parallel q_1).
\end{equation}

From the experimental point of view, the leptonic decay of the $W$ boson is
the most interesting one. In a sequel to this paper we shall calculate the
corresponding NLO electroweak corrections to the decay 
$W^+(\uparrow)\to\ell^+\,\nu_\ell$.

As a further example of much topical interest we have discussed the Higgs 
decay modes $H\to W^-+W^{\ast+}(\to c\bar b,c \bar s)$ and
$H\to Z+Z^\ast(\to b\bar b,\,c\bar c)$ involving the off-shell $W^{\ast+}$
and $Z^\ast$ bosons. We find that quark-mass effects and scalar contributions
affect the rate and the angular decay distributions in these decays in a
non-negligible way especially in the vicinity of the threshold region. Quark
mass effects are also non-negligible for the overall rate. For example,
nonzero quark masses induce a scalar contribution to the rate which makes up
$8.6\,\%$ of the total rate for $H\to Z+Z^\ast(\to b\bar b)$.

It would be worthwhile to exploit the knowledge about charged and neutral
current spectral functions in the heavy quark sector which has been
accumulating over the last few decades for a precision analysis of the rates
of the decays $H\to W^-+W^{\ast+}(\to b\bar c)$ and
$H\to Z+Z^\ast(\to b\bar b,\,c\bar c)$.

\subsection*{Acknowledgements}
This work was supported by the Estonian target financed project No.~0180056s09,
and by the Estonian Science Foundation under grant No.~8769. J.G.K.\ would
like to acknowledge useful discussions with B.~J\"ager and K.~Schilcher.
S.G.\ acknowledges the support by the Deutsche Forschungsgemeinschaft (DFG)
under Grant No.~436~EST~17/1/06 and by the Forschungszentrum of the
Johannes-Gutenberg-Universit\"at Mainz ``Elementarkr\"afte und Mathematische
Grundlagen (EMG)''.

\begin{appendix}

\section{Decay rate terms}
\setcounter{equation}{0}\def\theequation{A\arabic{equation}}
In this appendix we present analytical expressions for the polarized decay 
functions introduced in the main text. For the tree-graph contributions we 
define logarithmic decay rate terms
\begin{eqnarray}
\ell_1=\ln\pfrac{1+\mu_1-\mu_2+\sla}{1+\mu_1-\mu_2-\sla},&&
\ell_2=\ln\pfrac{1-\mu_1+\mu_2+\sla}{1-\mu_1+\mu_2-\sla},\nonumber\\[12pt]
\ell_0=\ln\pfrac{(1-\sqrt{\mu_1})^2}{\mu_2},&&
\ell_4=\ln\pfrac{(1+\sqrt{\mu_1})^2-\mu_2}{\sqrt{\mu_1}}\qquad
\end{eqnarray}
and the linear combination $\ell_3=\ell_1+\ell_2$. One further has 
dilogarithmic decay rate terms given by
\begin{eqnarray}
I^\ell_z(0)\!\!\!&=&\!\!\!\Li_2(-z_+)-\Li_2(-z_-)
  +\Li_2\pfrac{z_+-\sqrt{\mu_1}}{\sqrt{\mu_1}z_+-1}
  -\Li_2\pfrac{\sqrt{\mu_1}z_+-1}{z_+-\sqrt{\mu_1}},\\
S^\ell_z(0)\!\!\!&=&\!\!\!\Li_2\pfrac{1-\mu_1-\mu_2-\sla}{1-\mu_1-\mu_2+\sla}
  +\Li_2\pfrac{1-\mu_1+\mu_2-\sla}{1-\mu_1+\mu_2+\sla}
  +\Li_2\pfrac{1+\mu_1-\mu_2-\sla}{1+\mu_1-\mu_2+\sla}
  \strut\nonumber\\&&\strut
  -\frac{\pi^2}2+\frac12\ln^2\pfrac{1-\mu_1-\mu_2-\sla}{1-\mu_1-\mu_2+\sla}
  +\ln\pfrac\lambda{2\mu_1\mu_2}
  \ln\pfrac{1-\mu_1-\mu_2-\sla}{1-\mu_1-\mu_2+\sla}
  \strut\nonumber\\[3pt]&&\strut
  +2\ln(2\sqrt{\mu_1})\ln(2\sqrt{\mu_2})
  -2\ln(1-\mu_1+\mu_2+\sla)\ln(1+\mu_1-\mu_2+\sla),\\[3pt]
I^\ell_1(0)\!\!\!&=&\!\!\!\Li_2(\mu_1)-\Li_2(\sqrt{\mu_1}z_+)
  -\Li_2(\sqrt{\mu_1}z_-)-\frac{\pi^2}6\strut\nonumber\\&&\strut
  +\frac12\Li_2\pfrac{(z_--\sqrt{\mu_1})^2}{(1-\sqrt{\mu_1}z_-)^2}
  +\frac12\Li_2(z_-^2)
  -2\Li_2\pfrac{\sqrt{\mu_1}(\sqrt{\mu_1}-z_-)}{1-\sqrt{\mu_1}z_-}
  \strut\nonumber\\&&\strut
  +\ln\pfrac{1-z_-^2}{1-\mu_1}\ln\pfrac{z_--\sqrt{\mu_1}}{1-\sqrt{\mu_1}z_-}
  +\ln z_-\ln(z_+-z_-),\\
S^\ell_1(0)\!\!\!&=&\!\!\!\Li_2(z_-)-\Li_2(-z_-)-\frac{\pi^2}4
  +\ln z_-\ln\pfrac{1-z_-}{1+z_-}\strut\nonumber\\&&\strut
  -\Li_2\pfrac{(1+\sqrt{\mu_1})(1-z_-)}{(1-\sqrt{\mu_1})(1+z_-)}
  +\Li_2\left(-\frac{(1+\sqrt{\mu_1})(1-z_-)}{(1-\sqrt{\mu_1})(1+z_-)}\right),
  \\[7pt]
I^\ell(0)\!\!\!&=&\!\!\!\Li_2(\sqrt{\mu_1}z_+)+\Li_2(\sqrt{\mu_1}z_-)
  -2\Li_2(\sqrt{\mu_1})+\ln^2z_-\ =\ S^\ell(0)
\end{eqnarray}
where
\begin{equation}
z_+=\frac1{2\sqrt{\mu_1}}\left(1+\mu_1-\mu_2+\sla\right)=z_-^{-1}\,.
\end{equation}
The decay rate terms originating from the loop corrections read
\begin{eqnarray}
\ell_A&=&2\ln\lambda-3\ln\sqrt{\mu_1\mu_2},\\
\ell_B&=&\ln\pfrac{\mu_1}{\mu_2},\\[3pt]
t_A&=&\left(\ell_A-\ln\left(1-(\sqrt{\mu_1}-\sqrt{\mu_2})^2\right)\right)\ell_3
  \nonumber\\&&\strut
  +\Li_2(1-\alpha_+)-\Li_2(1-\alpha_-)-2\real L'(\mu_1,\mu_2)\label{tA}
\end{eqnarray}
where
\begin{equation}
\alpha_+=\frac{1-\mu_1-\mu_2+\sla}{1-\mu_1-\mu_2-\sla}=\alpha_-^{-1}\,.
\end{equation}
The complex function $L'(\mu_1,\mu_2)$ is given by
\begin{eqnarray}
\label{L2}
L'(\mu_1,\mu_2)&=&L(\tilde v)
  -\Li_2\pfrac{(\sqrt{\mu_1}-\sqrt{\mu_2})(\tilde v+1)}{2\sqrt{\mu_1}}
  +\Li_2\pfrac{(\sqrt{\mu_1}-\sqrt{\mu_2})(\tilde v-1)}{2\sqrt{\mu_2}}
  \nonumber\\&&\strut
  +\Li_2\pfrac{-(\sqrt{\mu_1}-\sqrt{\mu_2})(\tilde v-1)}{2\sqrt{\mu_1}}
  -\Li_2\pfrac{-(\sqrt{\mu_1}-\sqrt{\mu_2})(\tilde v+1)}{2\sqrt{\mu_2}}
  \nonumber\\&&\strut
  +\ln\pfrac{(\sqrt{\mu_1}+\sqrt{\mu_2})-(\sqrt{\mu_1}-\sqrt{\mu_2})\tilde v}
  {(\sqrt{\mu_1}+\sqrt{\mu_2})+(\sqrt{\mu_1}-\sqrt{\mu_2})\tilde v}
  \ln\pfrac{\sqrt{\mu_1}}{\sqrt{\mu_2}}
\end{eqnarray}
where
\begin{equation}
\label{L1}
L(\tilde v)=\Li_2\pfrac{2\tilde v}{1+\tilde v}
-\Li_2\pfrac{-2\tilde v}{1-\tilde v}
  +i\pi\ln\pfrac{1-\tilde v^2}{4\tilde v^2}-\pi^2
\end{equation} 
and where the velocity parameter $\tilde v$ has been defined in 
Eq.~(\ref{velo}). The dilogarithmic and double-logarithmic terms in
Eq.~(\ref{L2}) are real whereas $L(\tilde v)$ is a complex function with its
real part explicitly given in Eq.~(\ref{L1}). In the limit $\mu_1=\mu_2=\mu$
all dilogarithmic terms and the double-logarithmic term in Eq.~(\ref{L2})
vanish and one remains with the contribution of $L(v)$ where $v=\sqrt{1-4\mu}$
is the usual velocity of the quarks. Note that the term $L'(\mu_1,\mu_2)$ is a
generalization of the equal-mass term ($\mu_1=\mu_2=\mu$)
\begin{equation}
L(v)=\Li_2\pfrac{2v}{1+v}-\Li_2\pfrac{-2v}{1-v}
  +i\pi\ln\pfrac{1-v^2}{4v^2}-\pi^2
\end{equation}
appearing in $e^+e^-\to t\bar t$ (see e.g.\ Ref.~\cite{Groote:2008ux}).

\section{Decay rate terms in the high-energy limit}
\setcounter{equation}{0}\def\theequation{B\arabic{equation}}
In the high-energy or, equivalently, in the mass-zero limit one obtains
\begin{eqnarray}
\ell_0&\to&-\ln\mu_2,\nonumber\\[3pt]
\ell_1&\to&-\ln\mu_1,\nonumber\\[3pt]
\ell_2&\to&-\ln\mu_2,\nonumber\\[3pt]
\ell_3&\to&-\ln\mu_1-\ln\mu_2,\nonumber\\[3pt]
\ell_4&\to&-\frac12\ln\mu_1
\end{eqnarray}
using the expansion~(\ref{helimit}). Further one has
\begin{equation}
z_+\to\frac1{\sqrt{\mu_1}},\qquad
z_-\to\sqrt{\mu_1}
\end{equation}
or, more precisely, $\sqrt{\mu_1}z_+\to1-\mu_2$. Finally, in the tree-graph
case, one obtains
\begin{eqnarray}
I^\ell_z(0)&\to&-\frac{\pi^2}3-\frac14\ln^2\mu_1-\frac12\ln\mu_1\ln\mu_2
  -\frac12\ln^2\mu_2,\nonumber\\
S^\ell_z(0)&\to&-\frac{\pi^2}2-\frac12\ln\mu_1\ln\mu_2-\frac12\ln^2\mu_2,
  \nonumber\\
I^\ell(0)&\to&\frac{\pi^2}6+\frac12\ln^2\mu_1,\nonumber\\
S^\ell(0)&\to&\frac{\pi^2}6+\frac12\ln^2\mu_1,\nonumber\\
I^\ell_1(0)&\to&-\frac{\pi^2}3-\frac14\ln^2\mu_1,\nonumber\\
S^\ell_1(0)&\to&-\frac{\pi^2}2.
\end{eqnarray}
For the decay rate terms deriving from the loop corrections one has
\begin{eqnarray}
\ell_A&\to&
  -\frac32(\ln\mu_1+\ln\mu_2),\nonumber\\[3pt]
\ell_B&\to&\ln\mu_1-\ln\mu_2,\nonumber\\[3pt]
t_A&\to&\pi^2+\ln^2\mu_1+\ln\mu_1\ln\mu_2+\ln^2\mu_2.
\end{eqnarray}
Finally, one obtains $A_S\to 3/4$ and $A_I\to 3/4$.

\section{Decay rate terms close to threshold}
\setcounter{equation}{0}\def\theequation{C\arabic{equation}}
Close to threshold where $\sla\to 0$ one has $\ell_0,\ell_1,\ell_2,\ell_3\to 0$
while $\ell_4\to\ln 4$. Note, however, that $\ell_4$ is always multiplied with
$\lambda$ and, therefore, does not give any contribution in this limit. In
order to calculate the dilogarithmic decay rate terms in this limit, one has to
expand $\sla$ more carefully. To that end we define a small quantity 
$\kappa$ where $\kappa^2=(1-\sqrt{\mu_1}-\sqrt{\mu_2}\,)$. On expanding in
$\kappa$ one obtains
\begin{equation}\label{lthresh}
\sqrt{\lambda\left(1,\mu_1,(1-\sqrt{\mu_1}-\kappa^2)^2\right)}
  =\sqrt{8\mu_1(1-\sqrt{\mu_1})}\ \kappa+O(\kappa^3).
\end{equation}
Using the expansion~(\ref{lthresh}), one can verify that
$I^\ell_z(0),S^\ell_z(0),I^\ell_1(0),S^\ell_1(0),I^\ell(0)\to 0$. Finally, the
decay rate terms that originate from the loop corrections read
\begin{eqnarray}
\ell_A&\to&2\ln\lambda
  -3\ln\left(\sqrt{\mu_1}(1-\sqrt{\mu_1})\right),\nonumber\\
\ell_B&\to&\ln\pfrac{\mu_1}{(1-\sqrt{\mu_1})^2},\nonumber\\[7pt]
t_A&\to&2\pi^2.
\end{eqnarray}
The term $\ell_A$ appears to be singular at threshold when $\lambda\to 0$.
However, $\ell_A$ is multiplied with $\sla$ in Eqs.~(\ref{ASI}) or $\ell_3$ in
Eq.~(\ref{tA}). Therefore, one finds that $A_I$ and $A_S$ are finite,
\begin{equation}
A_I,A_S\to 2\pi^2\sqrt{\mu_1}(1-\sqrt{\mu_1}).
\end{equation}
Note that the chromodynamic Coulomb singularity at threshold proportional
to $\alpha_s$ manifests itself in the overall factor
\begin{equation}\label{coulomb}
N=\frac{\alpha_s}{\pi\sla}N_cC_Fq^2.
\end{equation}

\section{Comparison with spectral function results}
\setcounter{equation}{0}\def\theequation{D\arabic{equation}}
There have been claims and counterclaims in the literature about the
correctness of previous results on vector and axial-vector spectral
functions at $O(\alpha_s)$. The present calculation gives us the opportunity
to check on previous results in the literature. According to the decomposition
\begin{equation}
-g^{\mu\nu}=-g^{\mu\nu}+\frac{q^{\mu}q^{\nu}}{q^2}
-\frac{q^{\mu}q^{\nu}}{q^2}
\end{equation} 
we define the vector and axial-vector spectral functions 
($H^{VV(AA)}=H^{VV(AA)}_{\mu\nu}(-g^{\mu\nu})$)
\begin{equation}
H^{VV(AA)}=H_{U+L}^{VV(AA)}-H_S^{VV(AA)}.
\end{equation}
Following our previous work it is convenient to define the linear 
combinations (not to be confused with
the linear combinations $H_1$ and $H_2$ defined in Sec.~4) 
\begin{equation}\label{H12}
H_S^1=\frac12(H_S^{VV}+H_S^{AA}), \qquad
H_S^2=\frac12(H_S^{VV}-H_S^{AA})
\end{equation}
and, accordingly, for $H^{1,2}$ and $H_{U+L}^{1,2}$.
At the Born-term level we obtain
\begin{eqnarray}
H^1({\it Born\/})&=&4N_cq^2(1-\mu_1-\mu_2),\qquad
H^2({\it Born\/})=16N_cq^2\sqrt{\mu_1\mu_2}, \\
H_S^1({\it Born\/})&=&2N_cq^2(1-\mu_1-\mu_2-\lambda), \qquad
H_S^2({\it Born\/})=-4N_cq^2\sqrt{\mu_1\mu_2}, \\
H_{U+L}^1({\it Born\/})&=&6N_cq^2(1-\mu_1-\mu_2-\lambda/3), \qquad
H_{U+L}^2({\it Born\/})=12N_cq^2\sqrt{\mu_1\mu_2}.\qquad
\end{eqnarray}
The NLO corrections read 
\begin{eqnarray}
\lefteqn{H^1(\alpha_s)\ =\ N\Big[4(1-\mu_1-\mu_2)A_S
  +2\mu_1(1+\mu_1)\ell_1+2\mu_2(1+\mu_2)\ell_2
  \strut}\nonumber\\[6pt]&&\strut
  +\left((1-\mu_1-\mu_2-\lambda)\lambda-8\mu_1\mu_2\right)\ell_3
  -(\mu_1-\mu_2)\lambda\sla\ell_B
  -2(1+\mu_1+\mu_2-\lambda)\sla\Big],\nonumber\\[7pt]
\lefteqn{H^2(\alpha_s)\ =\ 4\sqrt{\mu_1\mu_2}N\Big[4A_S-(3-\mu_1-3\mu_2)\ell_1
  -(3-3\mu_1-\mu_2)\ell_2+6\sla\Big]}
\end{eqnarray}
and
\begin{eqnarray}
\lefteqn{H_S^1(\alpha_s)\ =\ \frac N2\Big[4(1-\mu_1-\mu_2-\lambda)A_S
  -2\mu_1(\mu_1-\mu_1^2+16\mu_2-\mu_1\mu_2-4\mu_2^2)\ell_1
  \strut}\nonumber\\[6pt]&&\strut
  -2\mu_2(16\mu_1-4\mu_1^2+\mu_2-\mu_1\mu_2-\mu_2^2)\ell_2
  -3\left((1-\mu_1-\mu_2-\lambda)\lambda-6\mu_1\mu_2\right)
  \ell_3\strut\nonumber\\[6pt]&&\strut
  +3(\mu_1-\mu_2)\lambda\sla\ell_B
  +6(1-\mu_1-\mu_2-\lambda+2\mu_1\mu_2)\sla\Big],\nonumber\\[7pt]
\lefteqn{H_S^2(\alpha_s)\ =\ \sqrt{\mu_1\mu_2}N\Big[-4A_S
  +(3-\mu_1-3\mu_2)\ell_1\strut}\nonumber\\[6pt]&&\strut
  +(3-3\mu_1-\mu_2)\ell_2-6\mu_1\mu_2\ell_3-3(2+\mu_1+\mu_2)\sla\Big].
\end{eqnarray}
Finally,
\begin{eqnarray}
\lefteqn{H_{U+L}^1(\alpha_s)\ =\ \frac N2\Big[
  4\left(3(1-\mu_1-\mu_2)-\lambda\right)A_S
  +2\mu_1(2+\mu_1+\mu_1^2-16\mu_2+\mu_1\mu_2+4\mu_2^2)\ell_1
  \strut}\nonumber\\[6pt]&&\strut
  +2\mu_2(2-16\mu_1+4\mu_1^2+\mu_2+\mu_1\mu_2+\mu_2^2)\ell_2
  -\left((1-\mu_1-\mu_2-\lambda)\lambda-2\mu_1\mu_2\right)\ell_3
  \strut\nonumber\\[7pt]&&\strut
  +(\mu_1-\mu_2)\lambda\sla\ell_B
  +2(1-5\mu_1-5\mu_2-\lambda+6\mu_1\mu_2)\sla\Big],\nonumber\\[7pt]
\lefteqn{H_{U+L}^2(\alpha_s)\ =\ 3\sqrt{\mu_1\mu_2}N\Big[4A_S
  -(3-\mu_1-3\mu_2)\ell_1\strut}\nonumber\\[6pt]&&\strut
  -(3-3\mu_1-\mu_2)\ell_2-2\mu_1\mu_2\ell_3+(6-\mu_1-\mu_2)\sla\Big].
\end{eqnarray}
The normalization factor $N$ has been defined in Eq.~(\ref{defN}).
The result on $H_{U+L}^1(\alpha_s)$ has been listed before in the form
$2 H_{U+L}^1(\alpha_s)=H_{U+L}(\alpha_s)$ in Eq.~({\ref{upluslals}}).

When comparing to previous results in the literature we want to remind the 
reader that one uses a different terminology for the spectral function results
in the QCD sum rule community. What is called ``longitudinal'' there is called
``scalar'' here and what is called ``transverse'' there we call
``transverse + longitudinal ($U+L$)''.

We find agreement with the results of Ref.~\cite{Djouadi:1993ss} which were
given in terms of the correlator functions $\imag \Pi_{L,T}^{+/-}$. These
are related to our rate functions by
\begin{eqnarray}
H_S^1(\alpha_s)=-\frac N\pi\imag\Pi_L^+(s),&&
H_S^2(\alpha_s)=-\frac N\pi\sqrt{\mu_1\mu_2}\imag\Pi_L^-(s),
  \nonumber\\
H_{U+L}^1(\alpha_s)=\frac{3N}\pi\imag\Pi_T^+(s),&&
H_{U+L}^2(\alpha_s)=\frac{3N}\pi\sqrt{\mu_1\mu_2}\imag\Pi_T^-(s).
\end{eqnarray}
We find also agreement with Ref.~\cite{Schilcher:1980kr}, where the relevant
relations are
\begin{eqnarray}
16N_cs\,\rho^{V/A}(s)&=&-\frac3{4\pi^2}\sla(H^1\pm H^2),
  \nonumber\\
16N_cs\,\rho^{V/A}_L(s)&=&\frac3{4\pi^2}\sla(H_S^1\pm H_S^2)\,.
\end{eqnarray}
Taking into account the correction mentioned in the note added to
Ref.~\cite{Schilcher:1980kr} as well as the erratum of
Ref.~\cite{Schilcher:1980kr}, we could not find the obvious mistakes in the
integrals $J_1$ and $J_2$ mentioned in Ref.~\cite{Djouadi:1993ss}.

We mention that the correlator functions in
Ref.~\cite{Djouadi:1993ss,Schilcher:1980kr} have been obtained by calculating
the absorptive parts of the pertinent two-loop contributions. The resulting
analytical expressions for the correlator functions are somewhat simpler than
our expressions. The mutual agreement was checked numerically. 
 
\section{$O(\alpha_s)$ results in terms of $VV$, $AA$,\\
 $V\!A$ and $A\,V$ contributions}
\setcounter{equation}{0}\def\theequation{E\arabic{equation}}
When treating the decay $W^+\to q_1\bar q_2$ we have assumed a SM coupling
form for the weak decay symbolically written as $(V-A)^{\mu}(V-A)^{\nu}=
V^{\mu}V^{\nu}+A^{\mu}A^{\nu}-V^{\mu}A^{\nu}-A^{\mu}V^{\nu}$. In the general
case when the relative weight of the vector and axial-vector current is not as
simple as in the SM charged current transitions (as e.g.\ in $Z\to q\bar q$
or in SM extensions of the charged current transitions), one wants to be able
to avail of the corresponding $O(\alpha_s)$ expressions written in terms of
their $VV$, $AA$, $V\!A$ and $A\,V$ contributions.

In this appendix we shall therefore collect all $O(\alpha_s)$ expressions for
the polarized decay functions in terms of their $VV$, $AA$, $V\!A$ and $A\,V$
components. Extending the notation of Eq.~(\ref{H12}) to
\begin{eqnarray}
H_\alpha^1=\frac12(H_\alpha^{VV}+H_\alpha^{AA}),&&
H_\alpha^2=\frac12(H_\alpha^{VV}-H_\alpha^{AA}),\nonumber\\
H_\alpha^3=\frac i2(H_\alpha^{V\!A}-H_\alpha^{A\,V}),&&
H_\alpha^4=\frac12(H_\alpha^{V\!A}+H_\alpha^{A\,V}),
\end{eqnarray}
where $\alpha$ is any of $U+L,U,L,F,S,tt,t0,0t,00,\pm\pm$ or $1,2,3$ of
Sec.~4, one obtains at LO
\begin{eqnarray}
H_1^1({\it Born\/})&=&2N_cq^2(1-\mu_1-\mu_2), \qquad 
H_1^2({\it Born\/})\ =\ 4N_cq^2\sqrt{\mu_1\mu_2}, \nonumber \\
H_2^3({\it Born\/})&=&0, \qquad
H_2^4({\it Born\/})\ =\ -2N_cq^2\sla, \nonumber \\
H_3^1({\it Born\/})&=&2N_cq^2\lambda, \qquad
H_3^2({\it Born\/})\ =\ 0, \nonumber \\
H_{tt}^1({\it Born\/})&=&2N_cq^2(1-\mu_1-\mu_2-\lambda), \qquad
H_{tt}^2({\it Born\/})\ =\ -4N_cq^2\sqrt{\mu_1\mu_2} \\
H_{t0}^1({\it Born\/})&=&-2N_cq^2(\mu_1-\mu_2)\sla
  \ =\ H_{0t}^1({\it Born\/}), \qquad
H_{t0}^2({\it Born\/})\ =\ 0\ =\ H_{0t}^2({\it Born\/}).\nonumber
\end{eqnarray}
Using
\begin{eqnarray}
H_{\pm\pm}^{VV}&=&H_1^1+H_1^2, \qquad
H_{\pm\pm}^{AA}\ =\ H_1^1-H_1^2, \nonumber \\
H_{\pm\pm}^{V\!A}&=&\pm(H_2^4-iH_2^3), \qquad
H_{\pm\pm}^{AV}\ =\ \pm(H_2^4+iH_2^3), \nonumber \\
H_{00}^{VV}&=&H_1^1-H_3^1+(H_1^2-H_3^2), \qquad
H_{00}^{AA}\ =\ H_1^1-H_3^1-(H_1^2-H_3^2),\nonumber \\
H_{tt}^{VV}&=&H_{tt}^1+H_{tt}^2, \qquad
H_{tt}^{AA}\ =\ H_{tt}^1-H_{tt}^2, \nonumber \\
H_{t0}^{VV}&=&H_{t0}^1+H_{t0}^2, \qquad
H_{t0}^{AA}\ =\ H_{t0}^1-H_{t0}^2, \nonumber \\
H_{0t}^{VV}&=&H_{0t}^1+H_{0t}^2, \qquad
H_{0t}^{AA}\ =\ H_{0t}^1-H_{t0}^2,
\end{eqnarray}
one obtains
\begin{eqnarray}
H^{VV}_{\pm\pm}({\it Born\/})&=&2N_cq^2(1-\mu_1-\mu_2+2\sqrt{\mu_1\mu_2}), 
\nonumber \\ 
H^{AA}_{\pm\pm}({\it Born\/})\ &=&2N_cq^2(1-\mu_1-\mu_2-2\sqrt{\mu_1\mu_2}),
\nonumber \\
H_{\pm\pm}^{V\!A}({\it Born\/})\ &=&\mp 2N_{c}q^2 \sqrt{\lambda},\nonumber \\
H^{VV}_{00}({\it Born\/})&=&2N_cq^2(1-\mu_1-\mu_2-\lambda+2\sqrt{\mu_1\mu_2}), 
\nonumber \\ 
H^{AA}_{00}({\it Born\/})&=&2N_cq^2(1-\mu_1-\mu_2-\lambda-2\sqrt{\mu_1\mu_2}), 
\nonumber \\ 
H_{tt}^{VV}({\it Born\/})&=&2N_cq^2(1-\mu_1-\mu_2-\lambda-2\sqrt{\mu_1\mu_2}),
\nonumber \\
H_{tt}^{AA}({\it Born\/})\ &=&
2N_cq^2(1-\mu_1-\mu_2-\lambda+2\sqrt{\mu_1\mu_2}),\nonumber \\
H_{t0}^{VV,AA}({\it Born\/})&=&-2N_cq^2(\mu_1-\mu_2)\sla
  \ =\ H_{0t}^{VV,AA}({\it Born\/}).\qquad
\end{eqnarray}
Note that the amplitudes $H_2^{1,2}$ do not contribute to the parity even
pieces of $H_{\pm\pm}^{VV/AA}$.

The non-vanishing $\alpha_s$ contributions are given by
\begin{eqnarray}
H_1^1(\alpha_s)&=&N\Big[2(1-\mu_1-\mu_2)A_S-2\mu_1(1+7\mu_1-\mu_2)I^\ell_1
  \nonumber\\&&
  -\sqrt{\mu_1}(1-12\mu_1-5\mu_1^2-2\mu_2+4\mu_1\mu_2+\mu_2^2)S^\ell_1
  \nonumber\\&&
  -\mu_1(6+4\mu_1-7\mu_2)\ell_1+\mu_2(2+3\mu_1)\ell_2\nonumber\\&&
  -4\mu_1\mu_2\ell_3-(1-11\mu_1+\mu_2)\sla\Big],\nonumber\\
H_1^2(\alpha_s)&=&N\sqrt{\mu_1\mu_2}\Big[4A_S+4\mu_1I^\ell_1
  -2\sqrt{\mu_1}(1+\mu_1-\mu_2)S^\ell_1\nonumber\\&&
  -3(1-\mu_1-\mu_2)\ell_1-3(1-\mu_1-\mu_2)\ell_2+3\sla\Big],\nonumber\\[24pt]
H_2^3(\alpha_s)&=&4N\pi\sqrt{\mu_1\mu_2}\sla,\nonumber\\[3pt]
H_2^4(\alpha_s)&=&\frac N2\Big[-4\sla A_I
  +4(1-3\mu_1-\mu_1^2-2\mu_2+\mu_2^2)I^\ell\nonumber\\&&
  -2(2-\mu_1-\mu_1^2+\mu_2+\mu_1\mu_2)\ell_0-8\lambda\ell_4\nonumber\\&&
  +4\sla(1+2\mu_1-\mu_2)\ell_1+2\sla(2+\mu_1+\mu_2)\ell_2\nonumber\\&&
  +(3+14\sqrt{\mu_1}-3\mu_1+3\mu_2)\left((1-\sqrt{\mu_1})^2-\mu_2\right)\Big],
  \\[24pt]
H_3^1(\alpha_s)&=&\frac N2\Big[4\lambda A_S-12\mu_1(1+7\mu_1-\mu_2)I^\ell_1
  \nonumber\\&&
  -6\sqrt{\mu_1}(1-12\mu_1-5\mu_1^2-2\mu_2+4\mu_1\mu_2+\mu_2^2)S^\ell_1
  \nonumber\\&&
  -2\mu_1(20+13\mu_1+\mu_1^2-24\mu_2+\mu_1\mu_2+4\mu_2^2)\ell_1\nonumber\\&&
  +2\mu_2(4+12\mu_1-4\mu_1^2-\mu_2-\mu_1\mu_2-\mu_2^2)\ell_2\nonumber\\&&
  +\lambda\left(\mu_1+\mu_2-(\mu_1-\mu_2)^2\right)\ell_3\nonumber\\&&
  -(\mu_1-\mu_2)\lambda\sla\ell_B-2(3-36\mu_1-\mu_1^2+8\mu_1\mu_2-\mu_2^2)\sla
  \Big],\nonumber\\
H_3^2(\alpha_s)&=&N\sqrt{\mu_1\mu_2}\Big[12\mu_1I^\ell_1
  -6\sqrt{\mu_1}(1+\mu_1-\mu_2)S^\ell_1\nonumber\\&&
  +6\mu_1(1+\mu_2)\ell_1+6\mu_2(1+\mu_1)\ell_2
  -3(3-\mu_1-\mu_2)\sla\Big],\\[24pt]
H_{t0/0t}^1(\alpha_s)&=&N\Big[-2(\mu_1-\mu_2)\sla A_I\nonumber\\&&
  +2(\mu_1-5\mu_1^2-\mu_1^3-\mu_2+\mu_1\mu_2+\mu_1^2\mu_2+2\mu_2^2
  +\mu_1\mu_2^2-\mu_2^3)I^\ell\nonumber\\&&
  -(3\mu_1-\mu_1^2-2\mu_1^3-\mu_2-4\mu_1\mu_2+7\mu_1^2\mu_2+\mu_2^2
  +\mu_1\mu_2^2)\ell_0-4(\mu_1-\mu_2)\lambda\ell_4\nonumber\\&&
  +3(\mu_1-\mu_2)^2\sla\ell_1
  +\left(4(\mu_1-\mu_2)+\mu_2(1-\mu_1+\mu_2)+(\mu_1-\mu_2)^3\right)
  \sla\ell_3\nonumber\\&&
  -(\mu_1-\mu_1^2+\mu_2+2\mu_1\mu_2-\mu_2^2)\lambda\ell_B
  \mp(\mu_1-\mu_2)\lambda\sla\pi
  -\left((1-\sqrt{\mu_1})^2-\mu_2\right)\nonumber\\&&\strut\times
  \left(5\mu_1-8\sqrt{\mu_1}\mu_1+2\mu_1^2-2\mu_2+2\sqrt{\mu_1}\mu_2
  -10\mu_1\mu_2+2\mu_2^2\right)\Big],\\[12pt]
H_{t0/0t}^2(\alpha_s)&=&\sqrt{\mu_1\mu_2}N\Big[4\mu_1I^\ell
  +2(1-\mu_1-\mu_2+3\mu_1\mu_2)\ell_0
  -(1+\mu_1-\mu_2)\sla\ell_3-\lambda\ell_B\nonumber\\&&
  \pm4(\mu_1-\mu_2)\sla\pi+3\left((1-\sqrt{\mu_1})^2-\mu_2\right)
  \left(1-2\sqrt{\mu_1}-\mu_1-\mu_2\right)\Big].
\end{eqnarray}
The overall normalization factor $N$ has been defined in Eq.~(\ref{defN}).
Close to threshold $\sqrt{q^2}=m_1+m_2$ the $O(\alpha_s)$ results are given by
\begin{eqnarray}
H_{\pm\pm}^1&=&H_{\pm\pm}^2\ =\ H_{00}^1\ =\ H_{00}^2
  \ =\ 4N_cq^2\Bigg\{\sqrt{\mu_1\mu_2}\nonumber\\&&\strut
  +\frac{\alpha_s}{3\pi}\left(\frac{8\pi^2}\sla\mu_1\mu_2-\sqrt{\mu_1\mu_2}
  \Big(16-3(\sqrt{\mu_1}-\sqrt{\mu_2})(\ln\mu_1-\ln\mu_2)\Big)+O(\sla)\right)
  \Bigg\},\nonumber\\
H_{\pm\pm}^3&=&4N_cq^2\left\{\pm\frac{4\pi\alpha_s}{3\pi}
  \sqrt{\mu_1\mu_2}+O(\sla)\right\},\nonumber\\
H_{\pm\pm}^4&=&4N_cq^2\left\{\mp\frac{4\pi^2\alpha_s}{3\pi}
  \sqrt{\mu_1\mu_2}+O(\sla)\right\},\nonumber\\
H_{0t}^1&=&H_{t0}^1\ =\ 4N_cq^2\left\{-\frac{4\pi^2\alpha_s}{3\pi}
  \sqrt{\mu_1\mu_2}(\sqrt{\mu_1}-\sqrt{\mu_2})+O(\sla)\right\},\nonumber\\
H_{0t}^2&=&-H_{t0}^2\ =\ 4N_cq^2\left\{-\frac{4\pi\alpha_s}{3\pi}
  \sqrt{\mu_1\mu_2}(\sqrt{\mu_1}-\sqrt{\mu_2})+O(\sla)\right\},\nonumber\\
H_{tt}^1&=&-H_{tt}^2\ =\ 4N_cq^2\Bigg\{\sqrt{\mu_1\mu_2}\nonumber\\&&\strut
  +\frac{\alpha_s}{3\pi}\left(\frac{8\pi^2}\sla\mu_1\mu_2-\sqrt{\mu_1\mu_2}
  \Big(12-3(\sqrt{\mu_1}-\sqrt{\mu_2})(\ln\mu_1-\ln\mu_2)\Big)+O(\sla)\right)
  \Bigg\},\nonumber\\
\end{eqnarray}
where, again, identically vanishing contributions are not listed.
\end{appendix}

\end{document}